\pdfoutput=1
\RequirePackage{etex}
\documentclass[longauth]{aa}
\usepackage[utf8]{inputenc}
\usepackage{graphicx}
\usepackage{txfonts}
\usepackage{tabularx}
\usepackage{booktabs}
\usepackage{subcaption}
\usepackage{tikz}
\usepackage{mathtools}

\newcommand{\highlight}[1]

\usepackage{placeins}
\usepackage{lipsum} 
\usepackage{graphicx}
\usepackage{amsmath}
\usepackage{verbatim}
\usepackage{appendix}
\usepackage{hyperref}
\usepackage{ulem}
\usepackage{marvosym}
\usepackage{caption}
\usepackage{afterpage}
\usepackage[version=4]{mhchem}
\usepackage[switch]{lineno}
\usepackage[dvipsnames]{xcolor}
\usepackage{orcidlink}
\hypersetup{
    colorlinks=true,
    citecolor=blue,
    linkcolor=blue,
    urlcolor=red,
}

\begin{document} 

    \author{V. Vaulato}
    \authorrunning{Vaulato et al.} 
    
\title{Thermal emission spectroscopy of the ultra-hot Jupiter HAT-P-70b: NIRPS constraints on its refractory and volatile content}


\author{Valentina~Vaulato\inst{1,2}\thanks{Corresponding author:\\
\texttt{valentina.vaulato@unige.ch,\\
valentina.vaulato@utoulouse.fr}}
\and David~Ehrenreich\inst{1,3}
\and Stefan~Pelletier\inst{1}
\and Romain~Allart\inst{4}
\and Lina~Messamah\inst{1}
\and Melissa~J.~Hobson\inst{1}
\and Fr\'ed\'eric~Genest\inst{4}
\and Yolanda~G.~C.~Frensch\inst{5,1}
\and Christophe~Lovis\inst{1}
\and Lison~Malo\inst{4,6}
\and \'Etienne~Artigau\inst{4,6}
\and Fr\'ed\'erique~Baron\inst{4,6}
\and Susana~C.~C.~Barros\inst{5,7}
\and Bj\"orn~Benneke\inst{8,4}
\and Xavier~Bonfils\inst{9}
\and Fran\c{c}ois~Bouchy\inst{1}
\and Marta~Bryan\inst{10}
\and Bruno~L.~Canto~Martins\inst{11}
\and Ryan~Cloutier\inst{12}
\and Neil~J.~Cook\inst{4}
\and Nicolas~B.~Cowan\inst{13,14}
\and Jose~Renan~De~Medeiros\inst{11}
\and Xavier~Delfosse\inst{9}
\and Elisa~Delgado-Mena\inst{15,5}
\and Ren\'e~Doyon\inst{4,6}
\and Jonay~I.~Gonz\'alez Hern\'andez\inst{16,17}
\and David~Lafreni\`ere\inst{4}
\and Izan~de~Castro~Le\~ao\inst{11}
\and Claudio~Melo\inst{18}
\and Lucile~Mignon\inst{1,9}
\and Christoph~Mordasini\inst{19}
\and Francesco~Pepe\inst{1}
\and Rafael~Rebolo\inst{16,17,20}
\and Jason~Rowe\inst{21}
\and Nuno~C.~Santos\inst{5,7}
\and Damien~S\'egransan\inst{1}
\and Alejandro~Su\'arez~Mascare\~no\inst{16,17}
\and St\'ephane~Udry\inst{1}
\and Diana~Valencia\inst{10}
\and Rillck~Guilherme~de~Souza~Barros~de~Amorim\inst{11}
\and Charles~Cadieux\inst{1,4}
\and Eduardo~Cristo\inst{5,7}
\and Emily~Deibert\inst{22}
\and Dasaev~O.~Fontinele\inst{11}
\and Thierry~Forveille\inst{9}
\and Khaled~Al Moulla\inst{5,1}
\and Nicola~Nari\inst{23,16,17}
\and Bennett~Neil~Skinner\inst{12,24}
\and Atanas~K.~Stefanov\inst{16,17}
\and Vincent~Yariv\inst{9}
\and Gregg~Wade\inst{25,26}
\and Joost~P.~Wardenier\inst{19,4}
\and Drew~Weisserman\inst{12}}

\institute{
\inst{1}Observatoire de Gen\`eve, D\'epartement d’Astronomie, Universit\'e de Gen\`eve, Chemin Pegasi 51, 1290 Versoix, Switzerland\\
\inst{2}Universit\'e de Toulouse, CNES, CNRS, IRAP, 14 avenue Edouard Belin, 31400 Toulouse, France\\
\inst{3}Centre Vie dans l’Univers, Facult\'e des sciences de l’Universit\'e de Gen\`eve, Quai Ernest-Ansermet 30, 1205 Geneva, Switzerland\\
\inst{4}Institut Trottier de recherche sur les exoplan\`etes, D\'epartement de Physique, Universit\'e de Montr\'eal, Montr\'eal, Qu\'ebec, Canada\\
\inst{5}Instituto de Astrof\'isica e Ci\^encias do Espa\c{c}o, Universidade do Porto, CAUP, Rua das Estrelas, 4150-762 Porto, Portugal\\
\inst{6}Observatoire du Mont-M\'egantic, Qu\'ebec, Canada\\
\inst{7}Departamento de F\'isica e Astronomia, Faculdade de Ci\^encias, Universidade do Porto, Rua do Campo Alegre, 4169-007 Porto, Portugal\\
\inst{8}Department of Earth, Planetary, and Space Sciences, University of California, Los Angeles, CA 90095, USA\\
\inst{9}Univ. Grenoble Alpes, CNRS, IPAG, 38000 Grenoble, France\\
\inst{10}Department of Physics, University of Toronto, Toronto, ON M5S 3H4, Canada\\
\inst{11}Departamento de F\'isica Te\'orica e Experimental, Universidade Federal do Rio Grande do Norte, Campus Universit\'ario, Natal, RN, 59072-970, Brazil\\
\inst{12}Department of Physics \& Astronomy, McMaster University, 1280 Main St W, Hamilton, ON, L8S 4L8, Canada\\
\inst{13}Department of Physics, McGill University, 3600 rue University, Montr\'eal, QC, H3A 2T8, Canada\\
\inst{14}Department of Earth \& Planetary Sciences, McGill University, 3450 rue University, Montr\'eal, QC, H3A 0E8, Canada\\
\inst{15}Centro de Astrobiolog\'ia (CAB), CSIC-INTA, Camino Bajo del Castillo s/n, 28692, Villanueva de la Ca\~nada, Madrid, Spain\\
\inst{16}Instituto de Astrof\'isica de Canarias (IAC), Calle V\'ia L\'actea s/n, 38205 La Laguna, Tenerife, Spain\\
\inst{17}Departamento de Astrof\'isica, Universidad de La Laguna (ULL), 38206 La Laguna, Tenerife, Spain\\
\inst{18}European Southern Observatory (ESO), Karl-Schwarzschild-Str. 2, 85748 Garching bei M\"unchen, Germany\\
\inst{19}Space Research and Planetary Sciences, Physics Institute, University of Bern, Gesellschaftsstrasse 6, 3012 Bern, Switzerland\\
\inst{20}Consejo Superior de Investigaciones Cient\'ificas (CSIC), 28006 Madrid, Spain\\
\inst{21}Bishop's University, Dept of Physics and Astronomy, Johnson-104E, 2600 College Street, Sherbrooke, QC, Canada, J1M 1Z7, Canada\\
\inst{22}Department of Physics and Astronomy, University of Waterloo, 200 University W, Waterloo, ON N2L 3G1, Canada\\
\inst{23}Light Bridges S.L., Observatorio del Teide, Carretera del Observatorio s/n, 38500 G\"uimar, Tenerife, Spain\\
\inst{24}Origins Institute, McMaster University, 1280 Main St W, Hamilton, ON, L8S 4L8, Canada\\
\inst{25}Department of Physics, Engineering Physics, and Astronomy, Queen’s University, 99 University Avenue, Kingston, ON K7L 3N6, Canada\\
\inst{26}Department of Physics and Space Science, Royal Military College of Canada, 13 General Crerar Cres., Kingston, ON K7P 2M3, Canada\\
}


   \date{Received 2 June, 2026; Accepted 11 November, 2026}

    \titlerunning{Thermal emission spectroscopy of the atmosphere of the ultra-hot Jupiter HAT-P-70b}
    \authorrunning{Vaulato et al.}

 
  \abstract
  {\textit{Context.} The chemical makeup of exoplanets carries key insights into their origins and internal structures. The extremely high temperatures on the dayside of ultra-hot Jupiters transform chemical species (e.g. refractory metals) into gas form and allow volatile molecules (e.g. water) to dissociate into their constituents. In contrast, cooler nightside conditions enable condensation (e.g. Fe) and cold-trapping (e.g. Ti), as temperatures drop sufficiently for compounds to transition out of the gas phase.

\textit{Aim.} We aim to characterise the dayside atmosphere of the ultra-hot Jupiter HAT-P-70b via thermal emission spectroscopy, constraining its refractory and volatile budget, and assessing thermal dissociation processes and cold-trapping. Particular emphasis is placed on the relative abundances of rock-forming elements (e.g. Fe, Si, and Mg) and the depletion or survival of species such as Ti and H$_2$O.

\textit{Methods.} We analyse secondary eclipse observations of HAT-P-70b acquired with the NIRPS spectrograph. From the resulting emission spectrum, we infer the planet's atmospheric thermal structure and composition using a well-mixed atmospheric retrieval framework.

\textit{Results.}
We detect emission signatures from Fe, Mg, V, Si, and OH, while no significant H$_2$O signal. The presence of OH alongside the absence of water indicates strong thermal dissociation, with water likely depleted by more than 90\% in the probed layers. The retrieved temperature profile shows a pronounced thermal inversion. Elemental abundance ratios (Fe/Si, Mg/Si, V/Si) are broadly consistent with solar-like composition, while titanium appears sub-solar, possibly due to partial condensation and cold-trapping on the nightside.

\textit{Conclusions.}
Our findings indicate that HAT-P-70b's dayside atmosphere is dominated by refractory species and strong thermal processes. The detection of OH and H$_2$O absence indicate efficient water dissociation, while sub-solar titanium suggests nightside cold-trapping. Refractory element ratios are consistent with solar values, supporting a link between planetary and stellar compositions.}


   \maketitle
%
\section{Introduction}
\label{sec:Introduction}
The chemistry of gas giant exoplanet atmospheres is governed by a fundamental division between refractory and volatile species.
Refractory elements--metals and silicates (also commonly referred to as rocky elements)--are characterised by a superior heat resistance and elevated condensation temperatures \citep{lodders_solar_2003}. 
In cold environments, only volatile species with a low condensation temperature (such as water--H$_2$O) can exist in gas phase, while refractory compounds require higher temperatures to vaporise and thus are commonly found in liquid and solid form. This condition is typical of planetesimals being accreted in planetary bodies, where the most abundant refractory species like iron (Fe), silicon (Si) and magnesium (Mg) are sequestered in solids. In hot environments, such as those of hot and ultra-hot Jupiters, the temperatures can exceed 2000\,K, causing the refractory metals to evaporate and exist in their atomic (or even ionised) gas phase, substantially contributing to the overall gaseous budget of the planet. At equilibrium temperatures above 2000\,K, the atmosphere of ultra-hot Jupiters is thus assumed to be close to chemical equilibrium, cloud-free (as condensation can be ``ruled out''), and populated by atoms and ions~\citep{kitzmann_peculiar_2018}. However, condensation can still occur on the colder nightside of the planet and in the terminator regions which are observed mostly in transit. Refractory elements with high condensation temperatures (e.g. titanium--Ti, scandium--Sc, and aluminum--Al) can rain-out on the nightside of the planet 
becoming cold-trapped at high pressures (deep atmospheric layers). The cold-trapping inhibits these species to mix and circulate back to upper (lower pressures) atmospheric layers~\citep[as in WASP-121b and WASP-76b;][]{hoeijmakers_mantis_2024, pelletier_vanadium_2023}, leading to apparent depletion relative to chemical equilibrium expectations. Indeed, Ti-bearing species (e.g.\ Ti, titanium oxide--TiO, and ionised titanium--Ti+) are often not detected in the atmospheres of the colder ultra-hot planets using the cross-correlation method (with a few notable exceptions that showcase the hottest temperatures, such as WASP-33b, KELT-9b, TOI-2109b, TOI-1518b, WASP-189b, and MASCARA-1b; \citeauthor{cont_atmospheric_2022} \citeyear{cont_atmospheric_2022}; \citeauthor{hoeijmakers_atomic_2018}\citeyear{hoeijmakers_atomic_2018}; \citeauthor{ simonnin_2025_TOI1518b_time}\citeyear{simonnin_2025_TOI1518b_time}; \citeauthor{pelletier_2026_Ti_WASP-189b}\citeyear{pelletier_2026_Ti_WASP-189b}; \citeauthor{guo_detection_2024}\citeyear{guo_detection_2024}). Atmospheric characterisation likewise tends to constrain ultra-hot gas giant atmospheres as being deficient in gaseous titanium \citep[e.g. WASP-121b and WASP-76b;][]{hoeijmakers_hot_2020, merritt_non-detection_2020, kesseli_search_2020, tabernero_espresso_2021, maguire_high-resolution_2023, gandhi_retrieval_2023, pelletier_vanadium_2023, hoeijmakers_mantis_2024, pelletier_2025_crires, pelletier_2026_Ti_W121b_JWST/NIRISS}. However, for planets with very high estimated nightside temperature ($T_\mathrm{nightside} \gtrsim 2000$\,K), titanium signatures are more readily detected (e.g.\ in KELT-9b), suggesting that cold-trapping becomes less and less efficient as the temperature increases. 

In contrast to Ti, chemically similar but less naturally abundant vanadium (V) is more ubiquitously detected in the atmospheres of ultra-hot Jupiters \citep{prinoth_time-resolved_2023, pelletier_vanadium_2023, vaulato+2025_wasp-121b}. However, their oxides (i.e.\ TiO and VO) are strong optical absorbers and thought to drive thermal inversions observed in hot Jupiters atmospheres \citep{fortney_unified_2008}. The temperature increase boosts the ionisation of metals such as Fe, sodium (Na), and potassium (K), which release free electrons in the atmosphere which, in turn, pair with atomic hydrogen (as the extremely hot atmospheres of gas giants transitions from H$_2$- to H-dominated) to form the hydride ion (H$^-$) which has bound-free interactions in the optical and near-infrared, and free-free transitions at longer wavelength \citep{parmentier_thermal_2018, arcangeli_h-_2018, vaulato+2025_hydride}. While TiO and VO are major opacity sources in the optical, H$^-$ becomes less and less absorbent in the near-infrared (up to 1.6~$\mu$m), but still prevents any detection of metals in the transmission spectrum of ultra-hot Jupiters \citep{vaulato+2025_hydride}.

In parallel, water, a key volatile and primary coolant of the upper atmosphere of hot and ultra-hot Jupiters, has been found to be significantly thermally dissociated on the hotter dayside \citep{parmentier_thermal_2018}. With H$_2$O being a major opacity source in the near-infrared \citep{gandhi_h-_2020}, its dissociation can strongly affect observed spectra. Equilibrium chemistry predicts that water molecules break down when temperatures exceed 2000\,K \citep{kitzmann_fastchem_2024}, releasing hydrogen and oxygen which recombine to form molecular hydroxyl (OH). Hydroxyl has been unambiguously detected in only a handful of hot and ultra-hot Jupiters so far: WASP-121b \citep{bazinet_quantifying_2025}, WASP-76b \citep{gandhi_revealing_2024}, WASP-18b \citep{brogi_roasting_2023}, WASP-189b \citep{sanchez_2025_Mg/Si}, WASP-33b \citep{cont_atmospheric_2022}, and WASP-12b. Among these, only two detections--those for WASP-76b with CARMENES \citep{quirrenbach_2016_CARMENES, landman_detection_2021} and WASP-121b with NIRPS \citep{Bouchy+2025}--were achieved using 4-meter-class telescopes, demonstrating that such detections are already accessible with modest-aperture facilities. 

Constraining the refractory and volatile content of ultra-hot Jupiter atmospheres provides key insights into both planet formation \citep{lothringer_new_2021, chachan_breaking_2023, lothringer_2026_library} and interior composition \citep{dorn_2015_can_we, sanchez_2025_Mg/Si}. The rock-forming elements Fe, Mg, and Si govern the interior mineralogy of rocky planets, and their relative abundances are generally assumed to mirror those of the host star. However, direct measurements of their relative ratios--Fe:Mg:Si--are accessible in the atmospheres of hot and ultra-hot Jupiters, where these species exist in atomic gas form due to the extremely high temperatures. Consequently, these gas giants provide ideal laboratories for testing planet interior models and for assessing the assumption that planetary bulk compositions reflect those of their host stars.\\

The atmosphere of the ultra-hot Jupiter HAT-P-70b, with an equilibrium temperature exceeding 2500 K, is likely shaped by chemical processes such as water photodissociation, titanium cold-trapping, and a gaseous refractory content (unlike normal hot Jupiters where refractories are not part of the gas phase) that makes it a valuable test-bed for planetary interior models. \cite{bello-arufe_mining_2022} analysed one single transit observation acquired with the high-resolution spectrograph HARPS-N \citep{Cosentino+2012}, and reported on a profusion of neutral and ionised chemical species: ionised calcium (Ca+), neutral and ionised chromium (Cr, Cr+), neutral and ionised iron (Fe, Fe+), neutral hydrogen (H), magnesium (Mg) , sodium (Na), and vanadium (V). They found tentative evidence for Ti+ (and Ca), but no detection of neutral Ti, suggesting that HAT-P-70b may reside in the titanium transition regime--i.e. sufficiently hot for partial titanium ionisation, but not hot enough for complete vaporisation or inefficient enough for full cold trapping. They also reported on observed blue-shifted signals (a few km~s$^{-1}$) indicative of atmospheric winds transporting material from the dayside toward the nightside hemisphere \citep[also confirmed by][]{langeveld_2025_HP70b}. Subsequently, \cite{gandhi_retrieval_2023} analysed the same single HARPS-N transit and retrieved Ti-bearing species to be underabundant with respect to model predictions, likely because of being cold-trapped in deeper atmospheric layers on the nightside of the planet. Using CARMENES \citep{quirrenbach_2016_CARMENES} transit transmission spectra, \cite{gan_2026_HATP70b_revisiting} confirmed previous detections of Fe, H$\alpha$, Na, and Ca+, reported on a new tentative detection of neutral potassium (K), and estimated an upper limit on the helium (He) triplet absorption. Recently, also \cite{guo+2026_HATP70b} reported on thermal dayside observations acquired with both CARMENES and PEPSI \citep{strassmeier_2015_PEPSI, stassmeier_2018b_PEPSI} high-resolution spectrographs. Via cross-correlation, they confirmed previous detections of Fe, Fe+, Cr, Mg, Ca+, and Ti, while reporting on new detections of Al, aluminium hydride (AlH), and manganese (Mn). Tentative signals include carbon (C), Ca, Na, sodium hydride (NaH), and nickel (Ni). Their atmospheric retrieval constrained an overall solar-like composition, with the exception of Ti, V and Ca, interpreted as to be cold-trapped on the nightside of HAT-P-70b. They also retrieved a Ni-enriched atmosphere, potentially owing to the planet accreting Ni-rich planetesimals during its formation. More recently, \cite{sun2026_hatp70b_maroonxconstraining} presented two optical transit time series observed with MAROON-X \citep{seifahrt_development_2016, seifahrt_maroon-x_2018, seifahrt_-sky_2020}, reporting on a plethora of detected species. Moreover, they found the atmosphere of the planet to be Ni-enriched, confirming previous findings by \cite{guo+2026_HATP70b}. Their retrieval framework constrained the abundances of species identified via cross-correlation, but relies strongly on model assumptions when inferring the abundances of ionisable species such as Ca and Ti. However, they concluded that Ti is always found underabundant likely due to cold-trapping \citep[also consistent with][]{guo+2026_HATP70b}.
In this work, we analyse unreleased near-infrared NIRPS thermal dayside observations of the atmosphere of HAT-P-70b, with the aim to constrain its refractory and volatile budget. The paper is organised as follow. In Sect. 2, we describe the observation strategy. Section 3 outlines the data reduction and processing. Section 4 describes the forward modelling and the cross-correlation analysis. In Sect. 5, we detail the forward modelling and cross-correlation analysis. Section 6 focuses on the atmospheric retrieval recipe. In Sect. 7, we present the results and discuss their interpretation.


\section{Observations}
\label{sec:Observations}
\begin{figure*}[!h]
   \centering
\includegraphics[width=\columnwidth]{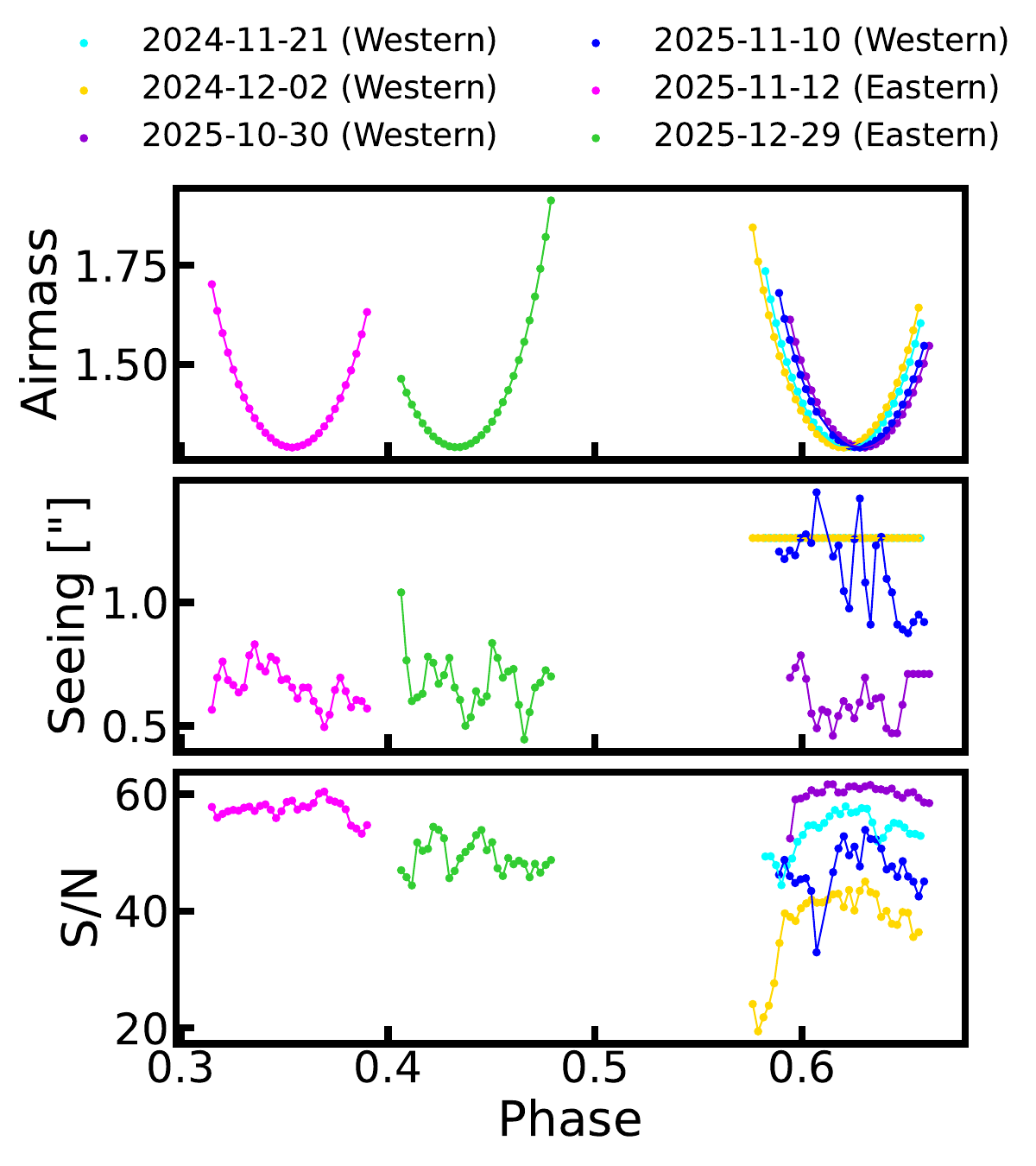}
\includegraphics[width=\columnwidth]{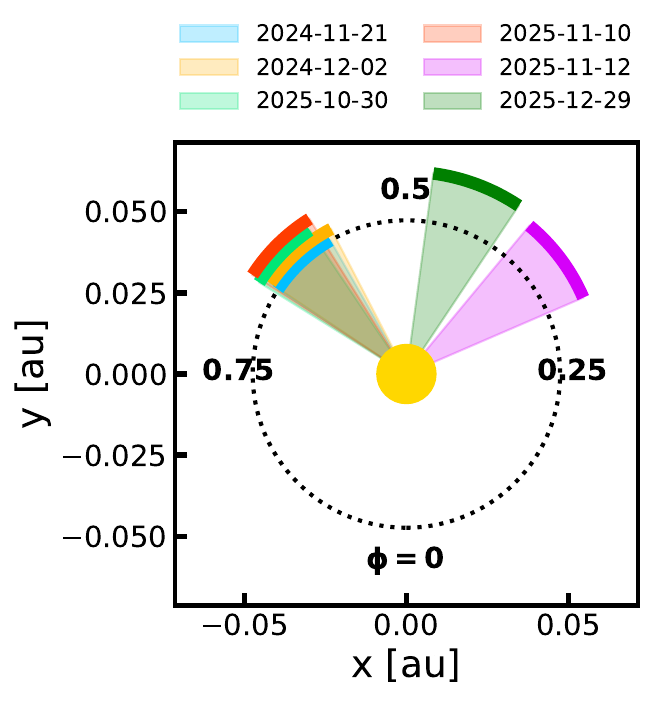}
   \caption{Left panel: NIRPS observing conditions as a function of orbital phase of the ultra-hot Jupiter HAT-P-70b. Different colours denote different observing nights. The top and middle panels show the evolution of the airmass and the average atmospheric DIMM seeing (i.e. the value measured by the Differential Image Motion Monitor) throughout each night, respectively. The DIMM seeing is averaged over each exposure, computed between the start and end times of the observation. The bottom panel displays the signal-to-noise ratio per pixel near 1.6 $\mu$m, extracted from the FITS header, as a function of phase.
   Right panel: Observed orbital coverage. This configuration illustrates a top-down view of the system, with the observer situated in the bottom direction. The observation campaign includes four post-eclipse (western dayside of the planet, $0.5<\phi<0.75$) and two pre-eclipse (eastern dayside, $0.25<\phi<0.5$) visits. On the axes is reported the distance from the star in astronomical units (au). Each observation segment is slightly offset for visualisation purposes.}
   \label{Figure:NightObservations}
\end{figure*}
We observed the dayside of the ultra-hot Jupiter HAT-P-70b~\cite[$R_p=1.87^{+0.15}_{-0.10}~R_\mathrm{Jup},~P=2.74432452^{+0.00000079}_{-0.00000068}~d,$ $T_{\mathrm{eq}}=2562^{+43}_{-52}\,\mathrm{K}$;][]{zhou_two_2019} with the Near-InfraRed Planet Searcher~\cite[NIRPS;][]{Bouchy+2017, Wildi+2022, artigau_nirps_2024, Bouchy+2025} spectrograph. NIRPS is the new high-resolution spectrograph mounted at the ESO 3.6-m telescope in La Silla, Chile. Designed to operate simultaneously with HARPS, it extends the wavelength coverage from the optical to $Y$, $J$, and $H$ photometric bands ($\lambda \in[0.98-1.80]~\mathrm{\mu m}$). The instrument is fibre-fed and assisted by adaptive optics, enabling high-precision radial velocity measurements and accurate atmospheric characterisation of exoplanets. The adaptive optics system provides NIRPS with a High Accuracy (HA) fibre covering a diameter of 0.4 arcsec on the sky for a greater spectral resolution (maximum of 96\,300 for the science fibre), and a larger High Efficiency (HE) fibre with a field of view of 0.9 arcsec. The NIRPS HE mode allows to reduce the impact of modal noise at the reddest wavelengths while keeping a good spectral resolution (maximum of 80\,300 for the science fibre) with respect to the HA mode. 

We observed the ultra-hot Jupiter HAT-P-70b as part of the NIRPS Guaranteed Time Observations 
(GTO), targeting orbital phases before and after the secondary eclipse for approximately 5 hours per observing window. The ultra-hot gas giant HAT-P-70b is included in the emission survey carried out within the Work Package 3 \citep[WP3;][]{Bouchy+2025, allart_2025_nirps_w69b} devoted to the atmospheric characterisation of exoplanets. We gathered a total of six visits: four post-eclipse (western dayside) on the nights of 2024 November 21, 2024 December 21, 2025 October 30, and 2025 November 10, and two pre-eclipse (eastern dayside) on 2025 November 12 and 2025 December 18. NIRPS observations are set up in the HE mode to reduce the effect of modal noise. We gathered a partial dayside on the night of 2025 October 30. The observation started about twenty minutes later than expected because the temperature of the instrument was above the safety threshold, and it was ended about thirty minutes before because the closure of the dome was anticipated. The analysis presented in this work is performed on all six thermal emission datasets gathered with NIRPS combined together.
A summary of the observations can be found in Fig.~\ref{Figure:NightObservations}.
In order to combine all observing nights together, we first built a common phase grid, then we interpolated each night on the generated phase grid with a one-dimensional linear interpolation function. The interpolated values are evaluated at each phase.


\section{Data reduction and processing}
\label{sec:DataProcessing}
Following~\cite{vaulato+2025_wasp-121b}, all spectra were reduced by the automated NIRPS 
Data Reduction Software (DRS version 3.3.12
), which returns the echelle-order merged (\texttt{S1D}) and echelle-order separated (\texttt{S2D}) spectra corrected for the Barycentric Earth Radial Velocity (BERV). The starting point of the atmospheric analysis are the 
NIRPS two-dimensional delivered by the pipeline as a matrix where each row is a given order. Specifically, we use the \texttt{S2D$\_$BLAZE$\_$A.fits} files, that are the science fibre (\texttt{A}) acquired spectra neither blaze nor telluric corrected. Therefore, the initial point of our data reduction steps is a time series of spectra in the barycentric rest frame: $f(\lambda,t)|_\mathrm{BERV}$. The spectra are then Doppler shifted to the stellar rest frame according to the stellar reflex motion. The pre-processed data undergo further reduction steps, hereby summarised:

\textit{Bad pixels removal.} We applied a fine-tuned 6$\sigma$-clipping to the flux variation order-by-order to flag and remove all outliers whose flux value is higher than the threshold. The clipping is performed by i) dividing each spectral order of each exposure by the median flux, and ii) dividing each spectral channel by the temporal flux. Thus, the outcome is a residual matrix where pixels deviating more than 6$\sigma$ from the median of the corresponding spectral channel are flagged as outliers and removed via interpolation of the neighbouring pixels. The final corrected flux matrix is constructed by multiplying the residual matrix by the median spectral and temporal fluxes.

\textit{Preparatory telluric masking.} Near-infrared ground-based spectra are crucially contaminated by Earth's atmosphere absorption (i.e.\ H$_2$O, CH$_4$, and CO$_2$ bands) and emission lines (i.e.\ OH). Therefore, we masked out all spectral regions heavily contaminated by telluric lines, namely 1.3450--1.4465 $\mu$m. The telluric OH emission lines are suppressed individually by routinely masking the peaks corresponding to the wavelength centroid tabulated in the linelist by~\cite{oliva_lines_2015}.

\textit{Blaze and continuum correction.} Since the DRS reduced spectra are in a non-blaze corrected format, we accounted for continuum or blaze variations alongside the observed time-series by applying a double filtering approach~\citep{gibson_detection_2020}. We aligned all spectra to the same continuum level by: i)~fitting a box-filter of width 51 pixels to each exposure, and ii)~applying a Gaussian convolution with a standard deviation of 100 pixels to smooth the box filter.

\textit{Residuals removal.} We applied a second-order polynomial fit on the median spectrum as in~\cite{brogi_retrieving_2019} to smooth out first-order telluric and stellar residuals of the cleaning. The best-fit spectrum (computed night by night) is then divided from each single exposure.

\textit{Principal component analysis (PCA).} We applied a PCA routine to remove telluric residuals. The PCA searches for coherent temporal patterns in the spectrum, models them, and reconstructs the observed time series. In~\cite{vaulato+2025_hydride} (see their Appendix E), the authors compared different telluric correction methods and the PCA approach proved to be the most efficient technique to clean the near-infrared spectra where heavy telluric H$_2$O and CO$_2$ broad bands contaminate the most. However, PCA is efficient for stationary telluric signals and large changes of the planetary radial velocity~\citep{cheverall_feasibility_2024}, thus limited to short period exoplanets. We decided to remove 5 principal components to best compromise on the cleanliness of the data while preserving the planetary signal.

The cleaned time-series spectra are retained in the stellar rest frame to avoid introducing additional interpolation errors. In this frame, the planetary signal appears as rapidly shifting emitting spectral features, which can be extracted by cross-correlating the processed spectra with a model template of the planet’s thermal emission.
\section{Forward modelling and cross-correlation analysis}
\label{sec:CrossCorrelationAnalysis}
Following~\cite{vaulato+2025_hydride, vaulato+2025_wasp-121b}, we generated a synthetic thermal emission spectrum of HAT-P-70b using \texttt{SCARLET}~\citep{benneke_atmospheric_2012, benneke_how_2013, benneke_strict_2015, benneke_water_2019, benneke_sub-neptune_2019, pelletier_where_2021}. \texttt{SCARLET} computes high-resolution line-by-line thermal emission spectral models for a given temperature structure, chemical composition, and cloud properties (Fig. \ref{Fig:SCARLET_models}). \texttt{SCARLET} includes atomic and molecular cross-sections calculated with \texttt{HELIOS-K}~\citep{grimm_helios-k_2015, grimm_helios-k_2021} and abundances from~\texttt{FastChem}~\citep{stock_fastchem_2018, stock_fastchem_2022}. Specifically, \texttt{SCARLET} models H$_2$-He collision-induced opacities, molecular cross-sections for H$_2$O~\citep{polyansky_exomol_2018}, OH~\citep{rothman_hitemp_2010}, atomic cross-sections from the VALD database~\citep{ryabchikova_major_2015} (for Fe, C, Na, Ti, V, Si, Sc, Ca, Ni, Cr, Mg, Mn, Al), and continuum cross-sections for H$^-$ \citep[both bound-free and free-free regimes][]{Gray_2021}.
The fiducial model was initially generated at resolution $\mathrm{R=250\,000}$, then degraded to $\mathrm{R=80\,000}$ (for NIRPS HE mode) by convoluting the model with a rotational kernel accounting for the planet's synchronous rotation, and a Gaussian kernel to match the instrumental resolution. We tested a set of models assuming a range of different metallicity (from 0.1 to 10 $\times$ Solar). However, the atmospheric model that maximises the S/N of the iron detection (which we assume as reference) corresponds to a solar metallicity ($\times$ 1 solar). In order to generate the forward atmospheric model, we adopted an inverted temperature-pressure structure assuming no heat redistribution.

We used the fiducial HAT-P-70b's atmospheric model to calculate cross-correlation functions~\citep[CCF;][]{snellen_orbital_2010} which served as an initial diagnostic to reveal the atmospheric composition of HAT-P-70b. We cross-correlated the model spectrum with the pre-processed and cleaned data. The synthetic spectrum is Doppler shifted from radial velocity $\mathrm{-400}$ to $\mathrm{+400}~\mathrm{km~s^{-1}}$ at steps of $v=2~\mathrm{km~s^{-1}}$,
\begin{equation}
   \label{eq:CCF_function}
     \mathcal{CCF}(v, t) = \sum_{i=0}^{N} \frac{\mathfrak{R}_{i,\lambda,t}~T_{i}(v)}{\sigma_i^2}~,
\end{equation}
where $\mathfrak{R}_{i,\lambda,t}$ is the time-series of $N$ emission spectra as function of time $t$ (i.e. exposure) and wavelength $\lambda$, $T_{i}(v)$ are the computed \texttt{SCARLET} template model spectra Doppler shifted to a velocity $v$, and $\sigma_i$ represents the uncertainties on the data \citep{gibson_detection_2020}, which weight the signal-to-noise ratio of each observing night (i.e. the signal-to-noise level of each pixel).
The CCF thus calculated generate the phase dependent, two-dimensional cross-correlation trail maps for each tested atomic and molecular species. In order to maximise the signal-to-noise ratio (S/N), the cross-correlation values are co-added along the trail which corresponds to the motion of the planet in the velocity space
\begin{equation}
rv~(t,K_\mathrm{p})=K_{\mathrm{p}}~\mathrm{sin(2\pi~\phi(t))}+V_{\mathrm{sys}}-V_{\mathrm{barycenter}},
\end{equation}
where $K_{\mathrm{p}}$ is the planetary orbital velocity semi-amplitude, $V_{\mathrm{sys}}$ is the systemic velocity, $V_{\mathrm{barycenter}}$ is the barycentric velocity which is equal to $0~\mathrm{km~s^{-1}}$ because our spectra are set in the BERV corrected rest frame (see Sec.~\ref{sec:DataProcessing}), and $\phi(t)$ is the planet's orbital phase. As the cross-correlation trail map computed for OH still exhibited clear residual OH emission features from the telluric correction, we further masked a velocity range centred at $10 \pm 20$ km~s$^{-1}$ (Fig. \ref{Figure:CCFTrailMaps}). This mask has been applied to all cross-correlation calculations, as residual OH emission contaminates the entire dataset.

We  phase-folded the CCF values for different orbital configurations in the velocity-velocity space as varying $V_{\mathrm{sys}}$ from $\mathrm{-100}$ to $\mathrm{+100}~\mathrm{km~s^{-1}}$ at steps of $1~\mathrm{km~s^{-1}}$, and $K_{\mathrm{p}}$ from $\mathrm{-300}$ to $\mathrm{+300}~\mathrm{km~s^{-1}}$ at steps of $1~\mathrm{km~s^{-1}}$. The resulting $K_{\mathrm{p}}-V_{\mathrm{sys}}$ maps~\citep{brogi_signature_2012} show the planetary signal if there is a match between the data and the model of the planet's atmosphere in the cross-correlation computation. Furthermore, we converted summed cross-correlation values into S/N estimates by dividing them by the standard deviation of all values excluding those near the expected location of the planetary signal, in order to extract the significance of the feature. To calculate the standard deviation of the $K_{\mathrm{p}}-V_{\mathrm{sys}}$ map, we excluded a 20~$\times$~20~$\mathrm{km~s^{-1}}$ boxy region centred on the expected planetary $K_{\mathrm{p}}$ and $V_{\mathrm{sys}}$. We also discarded $K_{\mathrm{p}}$ velocities below 20~$\mathrm{km~s^{-1}}$, since the map ``smooths out'' in this region, which would artificially decrease the overall standard deviation of the map, and hence signal-to-noise ratio of the detections. The results of the cross-correlation search for gaseous species in the atmosphere of HAT-P-70b are shown in Fig. \ref{Figure:CCFTrailMaps}, Fig.~\ref{Fig:CCF_maps},  quantitatively reported in Table~\ref{Table:GaussianFitting}, and discussed in Sect.~\ref{sec:CrossCorrelationResults}.


\section{Bayesian atmospheric retrieval}
\label{sec:AtmosphericRetrieval}
Based on the cross-correlations results, we run an atmospheric retrieval in order to constrain quantitatively the chemical composition of the atmosphere of the ultra-hot Jupiter HAT-P-70b. 
Here \texttt{SCARLET} is coupled with the \texttt{emcee} code as a sampler~\citep{foreman-mackey_emcee_2013} and leverages the high-resolution log-likelihood approach proposed by~\cite{gibson_detection_2020}:
\begin{equation}
    \label{Eq:CCF-to-LogL_Mapping_Gibson2020_elegant}
    \log(\mathcal{L})=-\frac{N}{2}\log\bigg(\frac{\chi^2}{N}\bigg),
\end{equation}
where $N$ is the number of data points, and $\chi^2$ are the residuals of the data. 

In order to model the vertical structure of the atmosphere, we use a temperature-pressure profile sampled by eight nodes, distributed at different altitudes (i.e. pressures) 
(Fig. \ref{Figure:TP_profile_retrieved}). We display only pressures between 1 bar and 10$^{-6}$ bar, as the very upper atmosphere is assumed to be isothermal (down to 10$^{-8}$ bar).
Each node is associated to a fitted temperature at a fixed pressure, spanning the verticality of the atmosphere. These nodes serve as control points, with the final temperature–pressure profile constructed through interpolation between them. We assumed the atmosphere as stratified into 40 layers. We choose a smoothing prior of $\sigma_\mathrm{smooth} = 500\,\mathrm{K\,dex^{-2}}$ to fit the temperature-pressure profile of arbitrary shape while penalizing unphysically steep temperature gradients~\citep{pelletier_where_2021}. Given the temperature-pressure profile, the relative amounts of H$_2$, He, and H are calculated by \texttt{FastChem}~\citep{stock_fastchem_2018, stock_fastchem_2022}.
We set the planetary mass to $\mathrm{M_p\sim1.66~M_\mathrm{Jup}}$, as computed by~\cite{gandhi_retrieval_2023} after retrieving the planetary surface gravity. 

We adopt the free atmospheric retrieval recipe, which does not make any assumptions of chemical process at play in the atmosphere and is thus ``free'' to explore the parameter space without being limited to chemically plausible scenarios. Therefore, we assume the volume mixing ratios of individual species to be well-mixed (i.e. constant with altitude). We assume an uniform prior distribution $\mathcal{U}(-12, 0)$ (in $\log$ scale) for all absolute abundances of interest. Given the high equilibrium temperature of HAT-P-70b ($T_\mathrm{eq}\simeq 2500$\,K), we treat the planetary atmosphere as being cloud-free with the continuum level likely instead set by H$^-$. Following~\cite{vaulato+2025_hydride} (see their Sect. 5.3), we let the retrieval free to fit for the abundance of both H$^-$ and the electron density (e$^-$) in order to assess the continuum opacity given by the hydride bound-free and free-free transitions. In the retrieval setup, we include opacities from Fe, Mg, V, Si, Ti, OH, H$_2$O, H$^-$ (bf), and H$^-$ (ff). As in~\cite{vaulato+2025_wasp-121b} (see their Eq. 2 in Sect. 4.2 and Eq. 3 in Sect. 5.2), we fit independently for their logarithmic volume mixing ratio (VMRs), defined as $\log\mathrm{VMR}_\mathcal{X} \coloneqq \log\mathcal{X}$. The constrained atmospheric parameters are summarised in Table~\ref{Table:retrieval_results best-fit}. Additionally, we retrieve the atmospheric temperature, the semi-amplitude orbital velocity of the planet, the systemic velocity, and the planet's equatorial velocity to account for the planet's rotation.


\section{Results and discussion}
\label{sec:Results}

\subsection{Cross-correlation search for gases}
\label{sec:CrossCorrelationResults}
\begin{figure*}[!h]
   \centering
  \includegraphics[width=\textwidth]{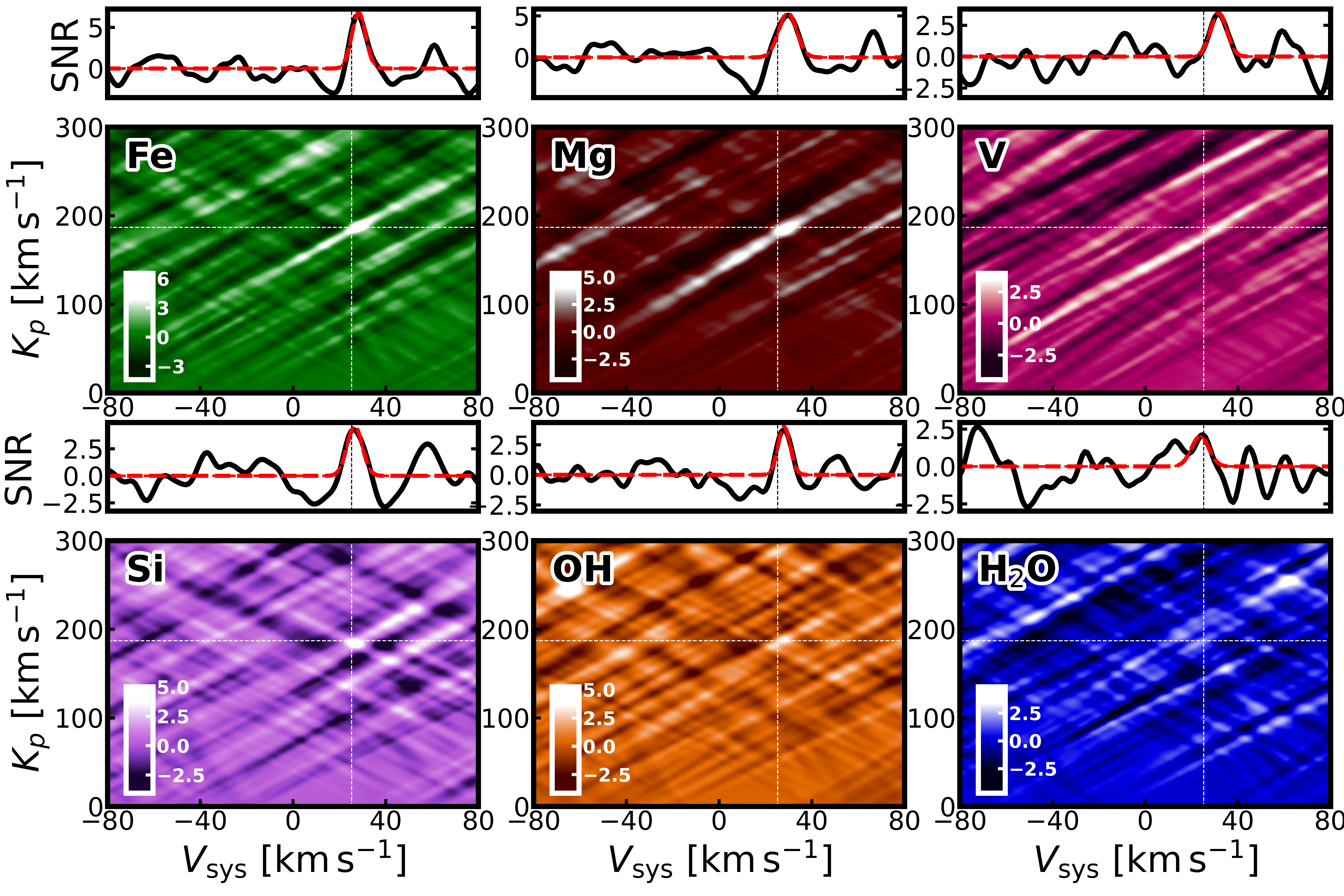}
   \caption{Detections of cross-correlation functions for Fe~(S/N~$=$~6.8), Mg~(S/N~$=$~5.4), V~(S/N~$=$~3.5), Si~(S/N~$=$~4.6) and OH~(S/N~$=$~4.0), and non-detection of H$_2$O (S/N~$\lesssim$~2.0) in the thermal emission spectrum of \hbox{HAT-P-70b}. Each squared panel shows the two-dimensional cross-correlation map in the velocity-velocity space ($\mathrm{K_p-V_{sys}}$). On top of each cross-correlation map is the one-dimensional cross-correlation function, expressed in terms of S/N as a function of systemic velocity, computed along a slice at the expected planetary $K_{\mathrm{p}}=187$ km~s$^{-1}$. The white dashed lines indicate the expected position of the planetary signal within the velocity-velocity space, namely the expected orbital~\citep[$\mathrm{\sim 187~km~s^{-1}}$,][]{bello-arufe_mining_2022} and systemic ~\citep[$\mathrm{\sim25.26~km~s^{-1}}$,][]{zhou_two_2019} velocities of \hbox{HAT-P-70b}. The rectangular panels at the top display the one-dimensional cross-correlation function for each species along the $\mathrm{K_p}$ axis, shown as signal-to-noise ratio versus systemic velocity.}
   \label{Fig:CCF_maps}
\end{figure*}
Using a cross correlation approach, we detect the spectral signatures of Fe (S/N = 6.8), Mg (S/N = 5.4), V (S/N = 3.5), Si (S/N = 4.6), and OH (S/N = 4.0) in the dayside atmosphere of HAT-P-70b by combining NIRPS near-infrared thermal dayside spectra obtained over six observational visits (see Sect.~\ref{sec:Observations}). However, with the sensitivity of our observations we did not detect a signal of H$_2$O (S/N $\lesssim$ 2).
The resulting cross-correlation $K_{\mathrm{p}}$–$V_{\mathrm{sys}}$ maps are shown in Fig.~\ref{Fig:CCF_maps}. Corresponding cross-correlation trail maps are displayed in Fig. \ref{Figure:CCFTrailMaps} in the Appendix. 

Each map is function of the signal-to-noise ratio, computed as the cross-correlation coefficients divided by the standard deviation evaluated over the full velocity–velocity space, excluding the region near the expected planetary signal (i.e. around the expected $K_{\mathrm{p}}$ and $V_{\mathrm{sys}}$).
\begin{table}[]
\centering
\caption{Gaussian fitting of detected signals.}
\label{Table:GaussianFitting}

\setlength{\tabcolsep}{3pt} 
\large

\begin{tabular}{lcccc}
\hline \hline
Element & $V_{\mathrm{sys}}$ [km\,s$^{-1}$] & FWHM [km\,s$^{-1}$] & Amplitude (S/N) \\
\hline
Fe   & 28.4 $\pm$ 3.0 & 7.0 & 6.8 \\
Mg   & 29.6 $\pm$ 3.8 & 9.0 & 5.4 \\
V    & 31.7 $\pm$ 3.5 & 8.2 & 3.5 \\
Si   & 26.6 $\pm$ 3.3 & 7.7 & 4.6 \\
OH   & 28.1 $\pm$ 2.6 & 6.2 & 4.0 \\
H$_2$O & 23.7 $\pm$ 3.7 & 8.8 & 2.0 \\
\hline
\end{tabular}

\end{table}
Above each map of Fig.~\ref{Fig:CCF_maps}, we present the one-dimensional cross-correlation function, expressed in terms of S/N as a function of systemic velocity, computed along a slice at the expected planetary $K_{\mathrm{p}}=187$ km~s$^{-1}$. We fit a Gaussian function to these profiles to estimate both the amplitude of the peak and its velocity position. The best-fit Gaussian parameters are reported in Table~\ref{Table:GaussianFitting}. 

We note that the signals from the refractory species (i.e. Fe, Mg, and V) do not peak at the expected location of the planetary signature in velocity space, it is rather ``elongated'' towards negative velocities, in contrast to the Si and OH signals. This behaviour is due to the differing contributions from the western and eastern daysides, as demonstrated for the refractory species iron, magnesium and silicon in Fig.~\ref{Figure:Fe_Western_Eastern_Contributions}, \ref{Figure:Mg_Western_Eastern_Contributions}, and \ref{Figure:Si_Western_Eastern_Contributions}, respectively.

Each detected chemical species exhibits a peak that is consistently slightly redshifted (in km~s$^{-1}$) relative to the reference value of $V_{\mathrm{sys}} = 25.26\pm0.11$ km~s$^{-1}$~\citep{zhou_two_2019}, in contrast to the more commonly observed blueshifted signals \citep{wardenier_modelling_2023, silva_espresso_2024, seidel+2025_w121b_dynamics_nature}.
Moreover, the projected effect of a day-to-night wind differs between transmission and thermal emission observations as they probe different pressure layers in the atmosphere. In transmission spectroscopy the observed signal mainly comes from the terminator region, while thermal emission spectroscopy probes the dayside hemisphere facing the observer. Therefore, the day-to-night wind pattern may produce different line-of-sight velocity shits in the two observing geometries.

A similar redshift was reported by \cite{yan_detection_2022} in the detection of CO on the thermal dayside of WASP-189b. They suggested that this shift could be explained by dayside to nightside winds \citep{tan_atmospheric_2019}. Another potential physical explanation to the observed shifts in systemic velocity may be due to an asymmetric day-to-night flow, in which the winds do not converge exactly at the anti-stellar point (longitude $\pm$180$^\circ$), but instead at a longitude offset from it. Hence, the net Doppler shifts induced by the wind circulation are different in pre-eclipse compared to post-eclipse (due to projection effects), which might lead to offsets in $V_{\mathrm{sys}}$ when fitting a sinusoidal function to the full planet's orbital trail. Alternatively, the redshifted signals we observe in Fig.~\ref{Fig:CCF_maps} may be a results of inaccuracies in the assumed reference systemic velocity. We therefore believe that the slight redshifts are most likely due to uncertainties in the determination of the assumed reference systemic velocity, rather than to a significant atmospheric circulation.

The dayside of ultra-hot Jupiters such as HAT-P-70b can reach extreme temperature which can break molecular constituents into their byproducts via thermal dissociation. This is the case of water that dissociates into atomic hydrogen and oxygen that further recombine to form molecular hydroxyl (OH)~\citep{parmentier_thermal_2018}. Molecular hydroxyl has been detected in the atmosphere of a few exoplanets. We report on the third existing detection of OH obtained with a 4-meter class telescope. We refer the reader to the introductory discussion in Sect.~\ref{sec:Introduction} for further details.

\subsubsection{Comparison with~\cite{guo+2026_HATP70b}}
Recently,~\cite{guo+2026_HATP70b} reported the detection of a wide range of chemical species (see summary in their Table 3) by combining four thermal dayside observations obtained with the optical PEPSI and the optical$+$near-infrared CARMENES spectrographs. They detect neutral aluminium (Al I), aluminium hydride (AlH), singly ionised calcium (Ca II), neutral chromium (Cr I), neutral iron (Fe I), singly ionised iron (Fe II), neutral magnesium (Mg I), neutral manganese (Mn I), and neutral titanium (Ti I), and report tentative detections of neutral carbon (C I), calcium (Ca I), sodium (Na I), sodium hydride (NaH), and neutral nickel (Ni I). With our observational campaign we confirm the detections of Fe and Mg, and we additionally report detections of V, Si, and OH. However, we do not detect any additional species reported in \cite{guo+2026_HATP70b}, for two reasons. First, spectral coverage: NIRPS observations span the near-infrared wavelength range from approximately 0.98 to 1.8~$\mu$m, thereby excluding spectral lines contributing in the optical. In contrast, the dataset utilised by~\cite{guo+2026_HATP70b} cover both the visible and the near-infrared wavelength ranges. Second, telescope facilities: while both NIRPS and CARMENES are mounted on 3.5-m class telescopes (with CARMENES located at the Calar Alto Observatory, Spain), PEPSI benefits from approximately twofold larger collecting area of the 2~$\times$~8.4-meter Large Binocular Telescope (Arizona, USA), thus significantly boosting the signal-to-noise ratio of the observations (between 134 and 233). 


\subsection{Free atmospheric retrieval}

\begin{table*}[!t]
    \centering
    \renewcommand{\arraystretch}{1.2}
    \caption{\hbox{HAT-P-70b} \texttt{SCARLET} atmospheric posteriors from the retrieval analysis of the combined NIRPS time series.}
    \begin{tabularx}{\textwidth}{
        >{\raggedright\arraybackslash}m{3cm}
        X
        >{\centering\arraybackslash}m{4cm}
        >{\centering\arraybackslash}m{4cm}
    }
        \toprule
        \toprule
        Parameters & Description & Value & Prior range \\
        \midrule

        $\log(\mathrm{Fe})$ & Log volume mixing ratio of iron
        & $-1.69^{+0.38}_{-0.54}$ & $\mathcal{U}(-12,0)^{a}$ \\

        $\log(\mathrm{Mg})$ & Log volume mixing ratio of magnesium
        & $-1.73^{+0.43}_{-0.57}$ & $\mathcal{U}(-12,0)^{a}$ \\

        $\log(\mathrm{V})$ & Log volume mixing ratio of vanadium
        & $-4.48^{+0.47}_{-0.81}$ & $\mathcal{U}(-12,0)^{a}$ \\

        $\log(\mathrm{Si})$ & Log volume mixing ratio of silicon
        & $-1.60^{+0.40}_{-0.53}$ & $\mathcal{U}(-12,0)^{a}$ \\

        $\log(\mathrm{Ti})$ & Log volume mixing ratio of titanium
        & $-4.5$ (3$\sigma$ upper limit) & $\mathcal{U}(-12,0)^{a}$ \\

        $\log(\mathrm{H}^{-})$ & Log volume mixing ratio of hydride
        & $-7.38$ (3$\sigma$ upper limit) & $\mathcal{U}(-12,0)^{a}$ \\

        $\log(\mathrm{e}^{-})$ & Log volume mixing ratio of electrons
        & Unconstrained & $\mathcal{U}(-12,0)^{a}$ \\

        $\log(\mathrm{H_2O})$ & Log volume mixing ratio of water
        & $-2.81$ (3$\sigma$ upper limit) & $\mathcal{U}(-12,0)^{a}$ \\

        $\log(\mathrm{OH})$ & Log volume mixing ratio of hydroxyl
        & $-2.18^{+0.46}_{-0.67}$ & $\mathcal{U}(-12,0)^{a}$ \\

        $T_1, T_2, \ldots, T_8$ [K] & Atmospheric temperature
        & See Fig.~\ref{Figure:TP_profile_retrieved}
        & $\mathcal{U}(100,6000)^{b}$ \\

        $K_{\mathrm{p}}$ [km\,s$^{-1}$] & Planetary velocity semi-amplitude
        & $187.29^{+0.76}_{-0.55}$ & $\mathcal{U}(156,276)$ \\

        $V_{\mathrm{sys}}$ [km\,s$^{-1}$] & Systemic velocity
        & $28.01^{+0.33}_{-0.26}$ & $\mathcal{U}(8,48)$ \\

        $v\sin i$ [km\,s$^{-1}$] & Planetary equatorial velocity
        & $4.72^{+0.40}_{-0.56}$ & $\mathcal{U}(0,20)$ \\

        \midrule

        $\log(\mathrm{Fe/Si})$ & Log Fe-to-Si abundance ratio
        & $-0.09^{+0.65}_{-0.67}$ & --- \\

        $\log(\mathrm{Mg/Si})$ & Log Mg-to-Si abundance ratio
        & $-0.13^{+0.68}_{-0.70}$ & --- \\

        $\log(\mathrm{V/Si})$ & Log V-to-Si abundance ratio
        & $-2.88^{+0.71}_{-0.90}$ & --- \\

        $\log(\mathrm{Ti/Si})$ & Log Ti-to-Si abundance ratio
        & $<-2.90$ & --- \\

        $\log(\mathrm{Mg/Fe})$ & Log Mg-to-Fe abundance ratio
        & $-0.04^{+0.69}_{-0.69}$ & --- \\

        $\log(\mathrm{V/Fe})$ & Log V-to-Fe abundance ratio
        & $-2.79^{+0.72}_{-0.89}$ & --- \\

        $\log(\mathrm{Ti/Fe})$ & Log Ti-to-Fe abundance ratio
        & $<-2.81$ & --- \\

        $\log(\mathrm{Si/Fe})$ & Log Si-to-Fe abundance ratio
        & $0.09^{+0.67}_{-0.65}$ & --- \\

        $\log(\mathrm{OH/H_2O})$ & Log OH-to-H$_2$O abundance ratio
        & $>0.63$ & --- \\

        \bottomrule
    \end{tabularx}

     \label{Table:retrieval_results best-fit}
     \begin{minipage}{\linewidth}
    \vspace{0.1cm}
    \small Notes: $^{a}$in $\log$ scale. $^{b}$With additional smoothing prior with $\sigma_\mathrm{smooth}=500\,\mathrm{K\,pressure~dex^{-2}}$.
    \end{minipage}
\end{table*}

\begin{figure*}
   \centering
   \includegraphics[width=1\textwidth, trim=25 0 5 0, clip]{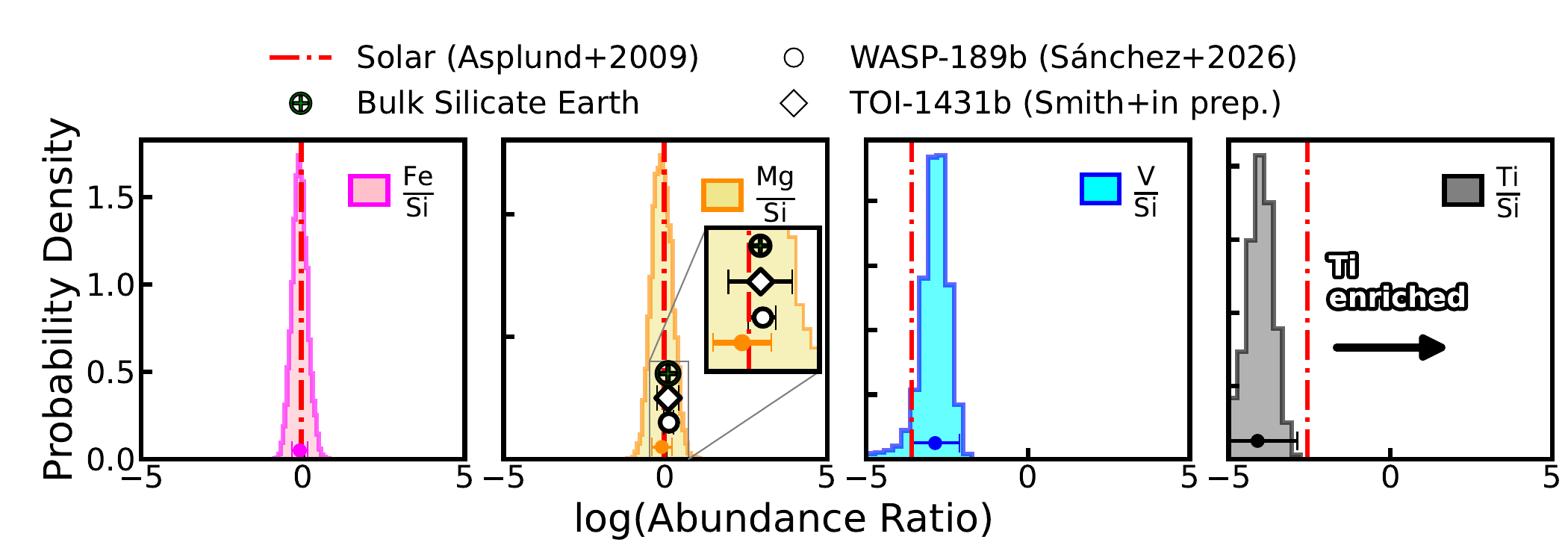}
   \caption{Retrieved elemental composition of the dayside atmosphere of HAT-P-70b from the free atmospheric retrieval. The panels from left to right show the probability density distributions of logarithmic abundance ratios of the refractory species relative to Si in the atmosphere of HAT-P-70b. Solar values \citep[red dotted-dashed line from][]{asplund_chemical_2009} are displayed as comparison. In the second panel from the left, the retrieved log(Mg/Si) ratio for WASP-189b \citep{sanchez_2025_Mg/Si}  and for TOI-1431b (P. Smith et al. in prep.) are overplotted for purposes of comparison, as well as an estimate of the bulk silicate composition of the Earth \citep{mcdonough_and_sun_1995_composition_earth}. Both the distribution of Fe and Mg relative to Si are consistent with a Solar composition within 1$\sigma$. On the contrary, the distribution of the ultra-refractory Ti relative to Si is sub-solar, likely due to the condensation of Ti-bearing species on the colder nightside hemisphere of HAT-P-70b.}
   \label{Figure:RelativeAbundances}
\end{figure*}

\begin{figure}[!h]
   \centering
\includegraphics[width=\columnwidth]{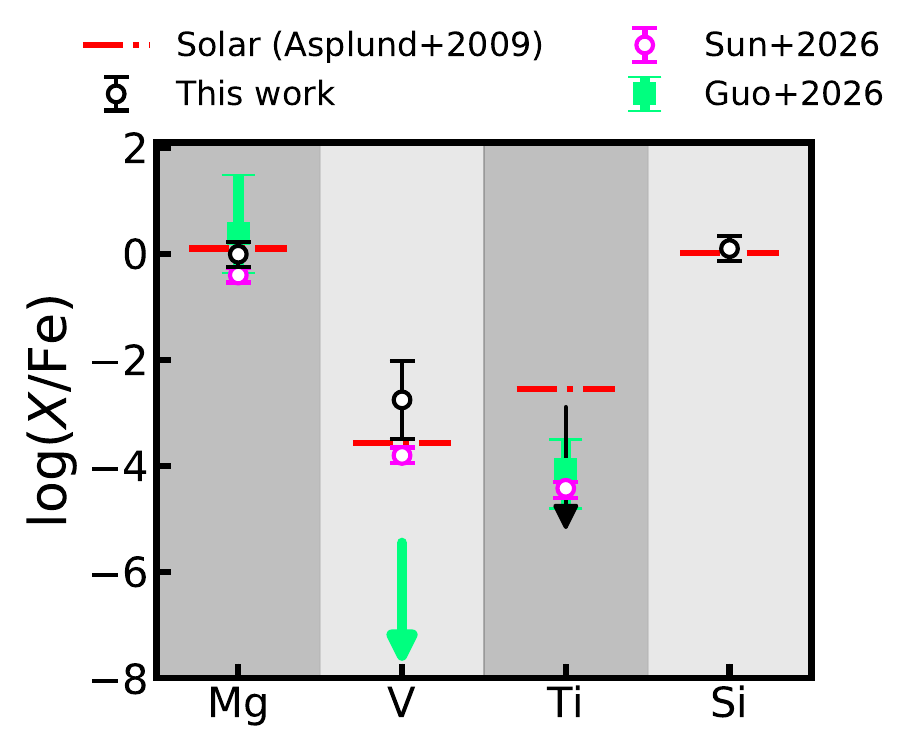}
   \caption{Retrieved elemental abundance of refractories relative to iron in logarithmic scale of the dayside atmosphere of HAT-P-70b (black dots) compared to \cite{guo+2026_HATP70b} (green) and Sun et al. accepted (pink). Solar values \citep[red dotted-dashed line from][]{asplund_chemical_2009} are displayed as comparison. 
   }
   \label{Figure:XoverFe}
\end{figure}

\begin{figure}[!h]
   \centering
\includegraphics[width=\columnwidth]{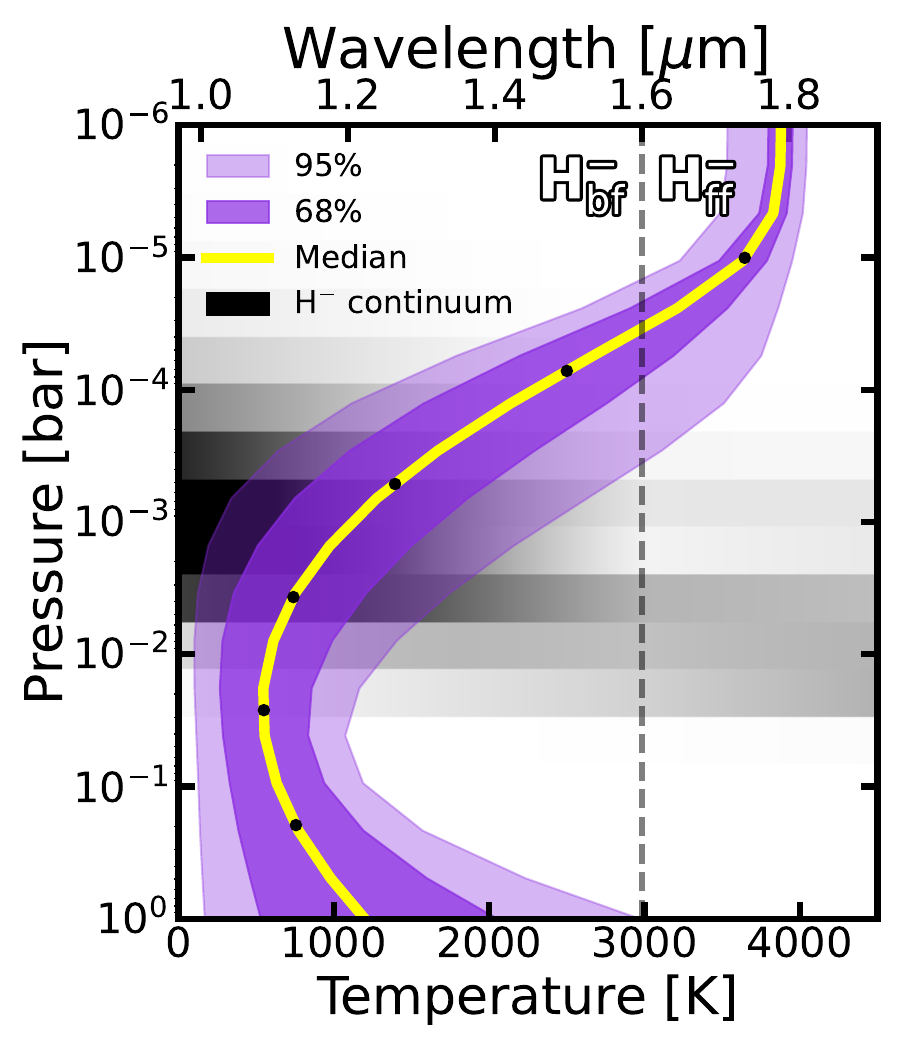}
   \caption{Retrieved vertical temperature-pressure structure of the dayside atmosphere of HAT-P-70b. The 68\% and 95\% confidence interval are shown as purple and violet contours, respectively. The median temperature-pressure profile is depicted as the yellow solid curve. The black dots represent the eight nodes distributed at different pressures to sample the vertical structure of the atmosphere between 1 bar and 10$^{-5}$ bar. The retrieved profile is coherent with the thermal inversions expected for the dayside atmospheres of ultra-hot gas giants \citep{fortney_unified_2008}. The gray shaded bands correspond to the two-dimensional contribution function (wavelengths vs. pressures). The contribution function shows the pressures probed by the continuum set by H$^{-}$ bound-free (H$^{-}_\textrm{bf}$) and free-free (H$^{-}_\textrm{ff}$) beyond 1.6 $\mu$m \citep{Gray_2021}.}
   \label{Figure:TP_profile_retrieved}
\end{figure}
Using our atmospheric retrieval framework, we quantitatively infer the chemical composition, vertical temperature-structure (temperature-pressure profile as function of altitude), and the velocity parameters of the dayside atmosphere of the ultra-hot Jupiter HAT-P-70b. 
The retrieved parameter posteriors of our setup free atmospheric retrieval (described in Sect.~\ref{sec:AtmosphericRetrieval}) are summarised in Table~\ref{Table:retrieval_results best-fit}. 

We recover a strong thermal inversion between 0.1 bar and 10$^{-4}$ bar, with a temperature difference of about 2000\,K (Fig.~\ref{Figure:TP_profile_retrieved}), consistent with the temperature-pressure profile retrieved in chemical equilibrium by \cite{guo+2026_HATP70b}. Unlike their free retrieval, we do not recover any double-inversion temperature profile (see their Fig. 8). \cite{gandhi_retrieval_2023} retrieved a uniform temperature–pressure profile (see their Fig. B.11) from transit transmission spectra of the planet, thereby probing the cooler terminator regions, where thermal inversions are not expected, in contrast to the hotter dayside region probed by our observations. We note that the retrieved temperature in the deep atmospheric layers (between 1 bar and 10$^{-2}$ bar) decreases significantly to roughly 1000\,K. While this scenario could arise if the planet has a remarkably high Bond albedo, we believe that this behaviour may due to either i) the retrieval being unable to accurately constrain the temperatures at the edges of the temperature-pressure profile, or ii) the prior imposed on the temperature-pressure profile being too tight. In the latter case, cooler temperatures at high pressures are preferred to minimise the second derivative penalty of the likelihood, namely the smoothing factor which we set to 500$\,\mathrm{K\,pressure~dex^{-2}}$ to enforce smoothness in the temperature-pressure profile. Consequently, the retrieval allows for lower temperatures deep down in the atmosphere (i.e. at higher pressures) to produce a smoother vertical temperature gradient. In order to demonstrate that the retrieval is unable to robustly constrain deep atmospheric pressure layers, we displays the two-dimensional contribution function in Fig. \ref{Figure:TP_profile_retrieved}. The contribution function set by the H$^-$ continuum (and its transitions at short and long infrared wavelengths) \citep{Gray_2021, vaulato+2025_hydride} shows that the atmosphere of HAT-P-70b is opaque at pressures deeper that 10$^{-2}$-10$^{-1}$ bar. However, even below the H$^-$ continuum, the temperature-pressure profile results quite constrained because it is driven by the prior on the temperature-pressure profile smoothing penalising the second derivative (i.e. the curvature).

From the atmospheric retrieval, we constrain the semi-amplitude of the orbital velocity of the planet, $K_{\mathrm{p}}=187.29^{+0.76}_{-0.55}$ km~s$^{-1}$, to be consistent with the reference value from~\cite{bello-arufe_mining_2022} \citep[i.e. $K_{\mathrm{p}}=187$ km~s$^{-1}$, which has been calculated according to the orbital parameters in][]{zhou_two_2019}, and the systemic velocity, $V_{\mathrm{sys}} = 28.01^{+0.33}_{-0.26}$ km~s$^{-1}$, which is slightly offset from our reference value, $V_{\mathrm{sys}} \simeq 25$ km~s$^{-1}$, from~\cite{zhou_two_2019}.
%

The retrieved rotational broadening parameter
results in a non-point-like signature in the $K_{\mathrm{p}}-V_{\mathrm{sys}}$ maps, which is diffused in the velocity-velocity space to a greater extent than expected accounting for the instrument resolution of NIRPS alone. This is most likely caused by the planet assumed to be a 3-dimensional rotating disc which we account for and model, with a planetary equatorial rotation velocity that is retrieved to be $4.72^{+0.40}_{-0.56}$ km~s$^{-1}$. The retrieved broadening is consistent with the calculated planet's rotation rate (i.e. the planet's equatorial velocity) $V_\mathrm{eq.~rot}=\omega\times R_\mathrm{p}= (2\pi/P)\times R_\mathrm{p}\simeq4.28$ km~s$^{-1}$, where both the period $P$ and the planetary radius $R_\mathrm{p}$ are derived by~\cite{zhou_two_2019}.

From the free retrieval, we obtain constraints for the volume mixing ratio of individual chemical species, as well as the relative abundance ratio. Relative abundance ratios are a more powerful diagnostic of the chemistry of a planet's atmosphere than individual abundances as they are less sensitive to model and data treatment assumptions~\citep{gibson_detection_2020, burningham+2017_retrieval_atmo_Ldwarfs}. 
We measure the ratios of $\log(\mathrm{Fe/Si})=-0.09^{+0.65}_{-0.67}$, $\log(\mathrm{Mg/Si})=-0.13^{+0.68}_{-0.70}$, and $\log(\mathrm{V/Si})=-2.88^{+0.71}_{-0.90}$ to be consistent with the equivalent elemental abundance ratios in the Solar photosphere. In contrast, we measure the ratios $\log(\mathrm{Ti/Si})<-2.90$ and $\log(\mathrm{Ti/Fe})<-2.81$ to be sub-solar. Our free retrieval also fails to constrain the individual abundance of both H$^-$ and the electron density. The reason is twofold: i) the spectral range of wavelengths covered by the observations is not sensitive to free-free transitions of hydrogen anions \citep[similar to NIRPS observations of both WASP-121b and WASP-189b][]{vaulato+2025_hydride,vaulato+2025_wasp-121b}; ii) the ionisation of metal species (which are the main donors of free electrons) is inefficient at the temperature and pressure probed by out data. The reference Solar abundance ratio is calculated based on the single-specie abundance inferred by~\cite{asplund_chemical_2009}. The probability density distributions of inferred relative abundance ratios are shown in Fig.~\ref{Figure:RelativeAbundances}. 

\subsubsection{log(Mg/Si)}
The most naturally abundant rock-forming refractory elements--Fe, Mg and Si--trace the interior mineralogy of rocky exoplanets as they reflect the bulk composition of disc solids. The internal structure of terrestrial planets consists in most Si and Mg present in the mantle, while Fe is partitioned between the mantle and the core \citep{spaargaren_2020_influence}. Since the interior composition of rocky planets is not yet observationally assessable, the models assume that the internal refractory content mirrors that of their host star, as both form from the same protoplanetary disc. Current theories of stellar formation posit that the elemental ratios in the stellar photosphere reflect those of the protostellar nebula from which both the star and the planetesimals (and thus the planet) formed \citep{thiabaud_2015_gas}. Consequently, their chemical composition are expected to be consistent, particularly with regard to the abundance ratios of ultra-refractory species, given their high condensation temperatures \citep{lodders_solar_2003}. 

The stellar elemental abundances of Fe, Mg, and Si are key constraints to reduce the degeneracy in interior structure models and to constrain mantle composition \citep{dorn_2015_can_we}. However, this assumption remains untested for the case of rocky planets. Assuming there are no deep radiative layers \citep{muller_can_2024, muller_ravit_2026_radiative_interior}, ultra-hot Jupiters, like HAT-P-70b, offer the chance to directly measure the abundance of these refractory species as they remain in their gas phase due to the very high temperature at play. 
To date, HAT-P-70b (this work), WASP-189b \citep{sanchez_2025_Mg/Si}, and TOI-1431b (P. Smith et al. in prep.) are the only three ultra-hot gas giants with measured Fe:Mg:Si ratios. Contrastingly to \cite{sanchez_2025_Mg/Si}, we cannot compare the retrieved abundance ratios Mg/Si and Fe/Si with the stellar values as the latter are not measured for HAT-P-70. We thus underline the difficulty in measuring stellar abundances, particularly in the case of A-type stars like HAT-P-70. We instead compare the relative abundances with the Solar composition and we retrieve it to be consistent with Solar value. All three results shown in the second panel of Fig. \ref{Figure:RelativeAbundances} \citep[i.e. this work, ][and P. Smith et al. in prep.]{sanchez_2025_Mg/Si} are consistent with Solar and with the Bulk Silicate Earth composition \citep[we assume log(Mg/Si)$_\oplus\simeq1.2$ from][]{mcdonough_and_sun_1995_composition_earth, chachan_breaking_2023}. 

Refractory elements span a range of volatility (i.e. how easily a species vaporises), and in the hotter regions of a protoplanetary disc the condensation sequence determines how much of each element is incorporated into solids at different temperatures \citep[][see their Fig. 1]{chachan_breaking_2023}. Partial condensation can therefore fractionate major elements in disc material, producing variations in refractory element ratios within forming protoplanets. Based on meteoritic evidence from our Solar System, the relative abundances of the most common refractory elements (i.e. Mg, Si, and Fe) are expected to vary by at most a few tens of percent, rather than by an order of magnitude.
Our results align with \cite{sanchez_2025_Mg/Si} and with P. Smith et al. (in prep.), supporting the hypothesis that the atmospheres of ultra-hot gas giants may preserve solar-like (stellar-like) rock-forming element ratios, which supports using stellar abundances as proxies in interior modelling.

It is also worth mentioning that the magnesium-to-silicon ratio can serve as a probe of silicate clouds formation--clouds composed of enstatite (MgSiO$_3$), forsterite (Mg$_2$SiO$_4$), and quartz (SiO$_2$). Indeed, in milder planet (warm to hot Jupiters), condensation is more likely than in ultra-hot Jupiters, favouring the formation of silicate clouds. 
\cite{calamari_2026_bridging_MgSiratio_brown_dwarfs} revisited the silicate cloud modeling of four hot Jupiters previously observed with JWST/MIRI \citep[WASP-17b, HD 189733b, WASP-39b, and WASP-107b;][]{dyrek_so2_2023, grant_jwst-tst_2023, inglis_quartz_2024, powell_sulfur_2024} in the context of brown dwarfs, vetting the hypothesis that giant planets accrete silicates from the protoplanetary disk with a Mg-to-Si ratio consistent with that of their host stars (see their Fig. 4).


\subsubsection{log(V/Si)}
Despite being consistent with Solar values within a 1$\sigma$ range, the peak of the posterior distribution of the $\log(\mathrm{V/Si})$ ratio results mildly super-solar. Our result lies between the vanadium abundances retrieved by \cite{sun2026_hatp70b_maroonxconstraining}, who analysed transmission spectroscopic time series gathered with MAROON-X. Their free retrieval finds the vanadium abundance (normalised to Solar) to be underabundant relative to the Solar value. In contrast, their chemical equilibrium retrieval yields a super-solar vanadium abundance. A similar result was reported by~\cite{smith_roasting_2024} (see their Fig. 8) for the thermal dayside of WASP-121b, where the atmosphere was found to be enriched in vanadium. 
Furthermore, we cannot compare our results with~\cite{guo+2026_HATP70b} as their dataset was not sensitive enough to claim any vanadium detection, therefore they were unable to constrain the abundance of V (neither the free nor the chemical equilibrium atmospheric retrieval), thus resulting in only an upper limit estimate (see their Fig. 9 and 10). 

\subsubsection{Where is Ti?}
Despite the lack of any Ti-bearing species detected in cross-correlation, we recovered a 3$\sigma$ upper limit on the Ti volume mixing ratio from the atmospheric free retrieval (see Table \ref{Table:retrieval_results best-fit}).
The retrieved under abundance of titanium relative to silicon and to iron compared to Solar (Fig. \ref{Figure:RelativeAbundances}, \ref{Figure:XoverFe}) may indicate that Ti is partially missing on the dayside atmosphere of HAT-P-70b. The condensation temperature of titanium ($\mathrm{T_{cond}}\sim1600~$K) is higher compared to other refractory metals such as Fe, Mg, or Si which are more ``resistant'' to condensation~\citep{lodders_solar_2003}. Therefore, it is easier for titanium to transition from the gaseous phase to a condensed state, especially in deep atmospheric layers (at high pressures) where the temperature decreases according to the inverted temperature-pressure profile (Fig.~\ref{Figure:TP_profile_retrieved}), and on the colder nightside refractory compounds can condense and rain-out \citep[e.g. Fe; ][]{ehrenreich_nightside_2020}. This suggests possible condensation of Ti on the planet's cooler nightside, which is dragged to the hotter dayside hemisphere via efficient atmospheric circulation resulting in a modest depletion of Ti-bearing species. Our result is consistent with the sub-solar Ti abundance found by~\cite{guo+2026_HATP70b}. This phenomenon is known as ``cold-trapping''~\citep{hoeijmakers_hot_2020,pelletier_vanadium_2023, hoeijmakers_mantis_2024}. However, we caution that our measurements may be biased due to TiO line list being inaccurate \citep{regt_quantitative_2022}. 
Ultra-hot Jupiters with $T_\mathrm{eq}\lesssim 2400~\mathrm{K}$ generally exhibit depleted Ti in their dayside atmospheres, whereas hotter planets ($T_\mathrm{eq}\gtrsim 2400~\mathrm{K}$) show detections of Ti-bearing species (e.g., Ti, Ti$+$, TiO) \citep{hoeijmakers_atomic_2018, prinoth_titanium_2022, prinoth_time-resolved_2023, prinoth+2025_W121b_4UT, pelletier_2026_Ti_W121b_JWST/NIRISS}. The ultra hot Jupiter HAT-P-70b ($T_\mathrm{eq}=2562~\mathrm{K}$) lies in a transition region between this two scenario. Indeed, both \cite{bello-arufe_mining_2022} and \cite{guo+2026_HATP70b} detected Ti$^+$ in the transmission spectrum of HAT-P-70b and Ti in its thermal emission spectrum, respectively. Also \cite{sun2026_hatp70b_maroonxconstraining} detect Ti in the transmission spectrum of HAT-P-70b. 
Both this work and \cite{sun2026_hatp70b_maroonxconstraining} retrieved (from the free retrieval approach) the abundance of titanium-to-iron to be $\approx2$ dex sub solar \citep[in agreement with the upper limit reported by][]{gandhi_retrieval_2023}, suggesting that titanium is only partially condensed and cold-trapped in deep atmospheric layers, as it has still been detected on both the terminator regions and the dayside hemisphere meaning that the circulation is efficient enough to transport Ti in gas phase. 
However, \cite{sun2026_hatp70b_maroonxconstraining} caution against the reliability of a well-mixed retrieval method used to quantify chemical abundances in transit transmission data. Indeed, their equilibrium chemistry retrieval yields a titanium-to-iron ratio of about 0.4 dex below Solar, so their atmospheric retrieval results are in disagreement. By definition, the free retrieval has no chemistry assumptions baked in; therefore, it does not account for ionisation processes which are expected to occur in the planet's atmosphere. In contrast, the chemical equilibrium retrieval takes these processes into account, and thus is expected to be more accurate, particularly when analysing nightside observations of a hot gas giant atmosphere.
The ultra-hot Jupiter WASP-121b also ($T_\mathrm{eq}\sim2360~\mathrm{K}$) exhibits deficiency in ultra-refractory titanium \citep{pelletier_2026_Ti_W121b_JWST/NIRISS} on its dayside atmosphere, as well as WASP-189b ($T_\mathrm{eq}\sim3350~\mathrm{K}$) whose abundance of Ti-to-Fe is found mildly under abundant on the thermal dayside (Pelletier et al. in review). In both cases, ionisation alone is insufficient to explain the depletion of Ti.

\subsubsection{Where is water?}
\begin{figure*}[h!]
    \centering
    \includegraphics[width=2.\columnwidth]{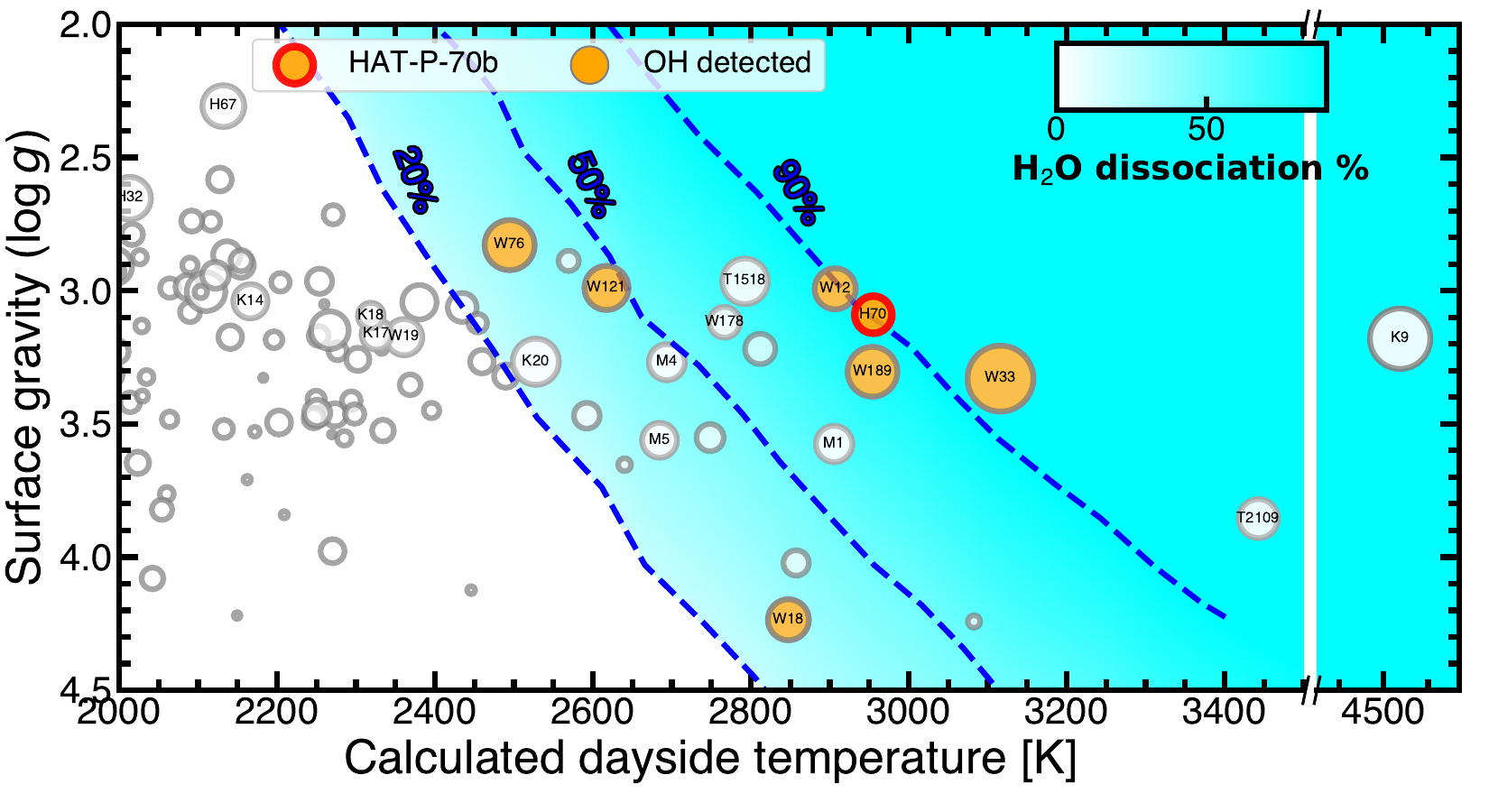}
    \caption{Existent OH detections from high-resolution observations shown as surface gravity vs. calculated dayside temperature compared to a portion of the exoplanet population. The dayside temperature is calculated according to the heat redistribution equation in \cite{parmentier_cloudy_2021} (see their Sect. 3.2). Planets for which OH has been detected are marked in orange. The size of the circular markers correspond to observational favourability. For the planet names, 'W'= WASP, 'K' = KELT, 'H' = HAT-P, 'T' = TOI, and 'M' = MASCARA. The red circle highlights HAT-P-70b within the population. The colorbar indicates the water dissociation percentage. The blue dashed lines denote \cite{parmentier_thermal_2018} models showing the percentage of thermally dissociated water.}
    \label{Figure:OHpopulation}
\end{figure*}
The dayside hemisphere of ultra-hot Jupiters can reach temperatures high enough for molecular compounds to thermally dissociate into their sub products. One example is water, which dissociates into oxygen and hydrogen which in turn combine to form OH. Specifically, the reaction chain is the following $\mathrm{H_2O} \rightleftharpoons 2\mathrm{H} + \mathrm{O}, \quad
\mathrm{H} + \mathrm{H} + \mathrm{X} \rightarrow \mathrm{H_2} + \mathrm{M}, \quad
\mathrm{O} + \mathrm{H_2} \rightarrow \mathrm{H} + \mathrm{OH}$, with $\mathrm{X}$ being any chemical species \citep{bazinet_quantifying_2025}. 
Water can dissociate via both equilibrium and disequilibrium chemical processes, depending on the altitude in the planetary atmosphere. At very low pressures (10$^{-5}$ bar, high altitudes) the photolysis (i.e.\ photodissociation) of H$_2$O via ultraviolet photon bombardment is prevalent. However, emission spectroscopy probes deeper atmospheric layers where thermal dissociation is the dominant mechanism to split water molecules into their byproducts.

Given the detection of OH, which provides evidence for the thermal dissociation of water, we aim to estimate the fraction of dissociated H$_2$O. We retrieve $\log(\mathrm{OH/H_2O})>0.63$ (Fig. \ref{Figure:OHoverH2O}), indicating that OH is more abundant than H$_2$O in the atmosphere of HAT-P-70b. This is further supported by the detection of hydroxyl alongside the lack of a strong H$_2$O signature. However, unlike other ultra-hot Jupiters \citep[WASP-18b, WASP-33b, and WASP-76b; respectively,][]{finnerty_keck_2023, brogi_roasting_2023, gandhi_revealing_2024}, we are unable to tightly constrain the relative abundance ratio, as only hydroxyl is detected while water remains undetected. However, the absence of a detection of water may also be due to a low S/N of the observations. To evaluate this possibility, we performed an injection-recovery test. The model is injected with a 1$\times$ nominal strength at the expected $K_{\mathrm{p}}$ and $V_{\mathrm{sys}}$. The test failed to recover the injected H$_2$O signal, indicating that--despite water is highly dissociated--the data sensitivity is not high enough to detect water. Indeed, higher signal-to-noise ratio data would be able to recover a water signal--IGRINS data for instance.

The relative abundance ratio leads to an estimated water dissociation percentage of $>90\%$ (Fig. \ref{Figure:OHpopulation}), in agreement with model predictions for the dayside 1.4~$\mu$m photosphere of HAT-P-70b \citep[][see their Equation 10 and Fig. 13]{parmentier_thermal_2018}.
However, at sub-millibar pressures, the abundances of H$_2$O and OH are expected to decrease \citep[e.g.\ see Fig. 6, left panel of][]{bazinet_quantifying_2025}. As a result, the strength of their spectral lines is reduced \citep{parmentier_thermal_2018}. Therefore, our assumption of a constant-with-altitude abundance may lead to an underestimation of the true abundances, as the model compensates by favouring lower values during the fit.


\section{Conclusions}
\label{sec:Conclusions}
In this work, we analysed thermal emission observations of the ultra-hot Jupiter HAT-P-70b in the near-infrared gathered with NIRPS. We collected six visits, including four post-eclipse and two pre-eclipse observations. Leveraging the high-resolution of the complete dataset, we constrained the chemical composition of the dayside atmosphere of the planet by combining the cross-correlation method and the Bayesian atmospheric retrieval framework.

We detected Si and OH for the first time in the atmosphere of HAT-P-70b, while no significant signature of H$_2$O is found. We also reported robust detections of Fe, Mg, and V. The simultaneous detection of OH and absence of water provides evidence for efficient thermal dissociation of H$_2$O in the probed atmospheric layers, with our retrieval suggesting that more than 90\% of water is dissociated. This result is consistent with theoretical expectations for ultra-hot atmospheres exceeding ~2000 K and confirms that OH can serve as a key tracer of water chemistry in such environments.

From a compositional perspective, the relative abundance ratios of the main rock-forming elements (Fe, Mg, Si) are consistent with solar values within uncertainties. This finding supports the hypothesis that giant planets preserve the bulk elemental ratios of their host stars, reinforcing the common assumption used in planetary interior modelling. In contrast, titanium was found to be significantly underabundant relative to solar, suggesting partial depletion likely caused by condensation and cold-trapping on the nightside, followed by inefficient recirculation to the dayside. This places HAT-P-70b in the transition regime where titanium is neither fully retained in the gas phase nor completely removed.
%
%

The retrieved temperature–pressure profile reveals a thermal inversion on the dayside, with temperature contrasts of the order of $\sim$2000 K across the photospheric region. However, our analysis excluded a double inversion pattern. This is consistent with expectations for ultra-hot Jupiters and supports the presence of strong near-infrared absorbers shaping the atmospheric structure.

This study demonstrates the capability of 4-meter class telescopes equipped with ultra-stable high-resolution spectrographs to probe the chemical and thermal properties of ultra-hot Jupiter atmospheres in emission. HAT-P-70b emerges as a valuable laboratory for studying the interplay between thermal dissociation, and condensation processes. 



\begin{acknowledgements}
Acknowledgments can be found in Appendix \ref{sec:Acknowledgments}.
\end{acknowledgements}


\bibliographystyle{aa} 
\bibliography{references}

\clearpage
\onecolumn
\pagebreak
\begin{appendix}

\section{Thermal emission spectroscopy}
\label{sec:EmissionSpectroscopy}
Once we cleaned the data, we calculated the thermal emission spectra solving the formal radiative transfer equation
\begin{equation}
    \frac{dI(\tau_{\lambda})}{d\tau_{\lambda}} = S(\tau_{\lambda}) - I(\tau_{\lambda}),
\end{equation}
where $I$ is the intensity at a specific wavelength $\lambda$, $\tau$ is the optical depth, and $S$ is the source function. For the emission-only case, when assuming local thermal equilibrium, one can set \hbox{$S(\tau_{\lambda})=B_{\lambda}(T)$}, where $B$ is the Planck function calculated at wavelength $\lambda$ and temperature $T$. The first order solution of the differential equation is written as the summation of the homogeneous and the particular solutions
\begin{equation}
    I(\tau_{\lambda})=I(\tau_{0,\lambda})~e^{\tau_{\lambda}-\tau_{0,\lambda}}-\int_{\tau_{0,\lambda}}^{\tau_{\lambda}}S(\tau'_{\lambda})~e^{\tau_{\lambda}-\tau'_{\lambda}}~d\tau'_{\lambda}~~.
\end{equation}
The emergent disc-integrated flux as function of wavelength is calculated over the full solid angle as
\begin{equation}
    F_{\lambda} = \oint  I(\tau_{\lambda})~cos(\theta)~dw = 2\pi\int_0^1I(\tau_{\lambda})~\mu_{\theta}~d\mu_{\theta}~,
\end{equation}
where $\mu_{\theta}=cos(\theta)$.
Transiting planets in circular orbits (tidally locked), will be eclipsed by their stars. As the planet approaches the secondary eclipse (phase $\phi = 0.5$, see Fig.~\ref{Figure:NightObservations}), the flux in a given wavelength bin is measured as
\begin{equation}
    \frac{\Delta F_{\lambda}}{F_{\lambda}}=\frac{F_{\mathrm{day},\lambda}~R_\mathrm{p}^2}{F_{\mathrm{day},\lambda}~R_\mathrm{p}^2 + F_{\star,\lambda}~R_\mathrm{\star}^2} \simeq \frac{F_{\mathrm{day},\lambda}~R_\mathrm{p}^2}{F_{\star,\lambda}~R_\mathrm{\star}^2}, 
\end{equation}
where $F_{\mathrm{day},\lambda}$ is the measured flux of the observed dayside of the planet, $R_\mathrm{p}$ is the planet's radius, while $F_{\star,\lambda}$ and $R_\mathrm{\star}$ are the stellar flux and radius, respectively.

In practice, the planetary emission spectrum is extracted as follow.
We shifted the corrected spectra (outcome of the data processing described in Sect.~\ref{sec:DataProcessing}) in the stellar rest frame and created an average stellar median spectrum containing the flux contribution from the star, while the planetary signal is too faint and undetectable as it is comparable with the noise
\begin{equation}
F_\lambda^{\rm median}|_\star =
\begin{cases}
F_{\lambda,\frac{N+1}{2}}|_{\star}, & N \text{ odd},\\[1ex]
\dfrac{F_{\lambda,\frac{N}{2}}|_{\star} + F_{\lambda,\frac{N}{2}+1}|_{\star}}{2}, & N \text{ even},
\end{cases}
\end{equation}
where $N$ is the number of time-series. Additionally, we interpolated all spectra to the same wavelength grid ($\tilde{F}_{\lambda,t}|_{\star}$) and subtracted the median stellar template to each time-series in the stellar rest frame, in order to isolate the planetary spectrum,
\begin{equation}
\mathfrak{R}_{\lambda,t} = \tilde{F}_{\lambda,t}|_{\star} - F_{\lambda}^{\rm median}|_\star, 
\quad t = 1, \dots, N.
\end{equation}

\clearpage
\section{Atmospheric forward models}
\begin{figure}[h!]
    \centering
    \includegraphics[width=0.6\linewidth]{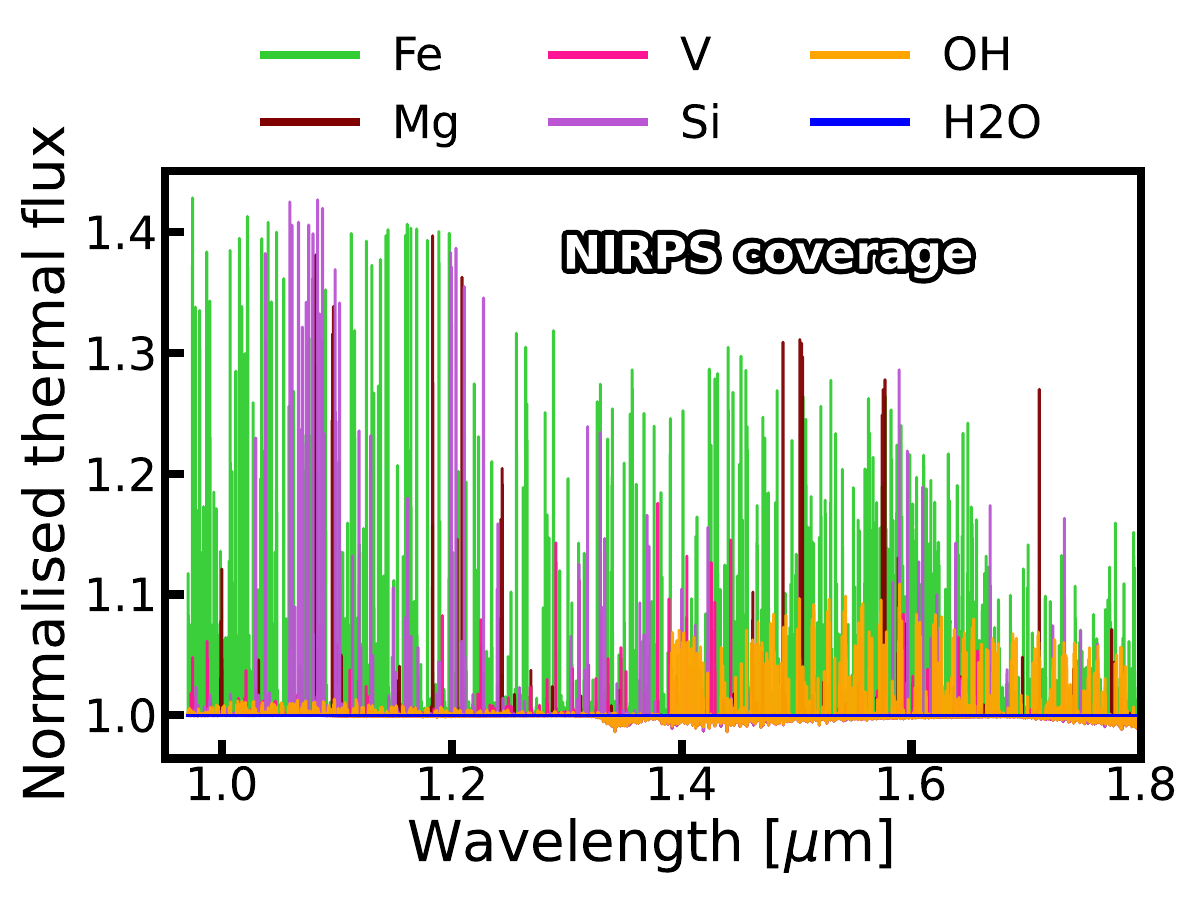}
    \caption{Atmospheric forward models (thermal flux as function of wavelength for Fe (green), Mg (brown), V (pink), Si (violet), OH (orange), and H$_2$O (blue) generated with the \texttt{SCARLET} framework over the range of wavelengths covered by NIRPS. The thermal flux is normalised by the continuum set by H$_2$O. We maintain consistent colour assignments for each chemical species as used in the cross-correlation blob maps shown in Fig. \ref{Fig:CCF_maps}.}
    \label{Fig:SCARLET_models}
\end{figure}

\section{Cross-correlation trail maps}
\clearpage
\begin{figure*}[h]
\centering
\includegraphics[width=0.47\textwidth]{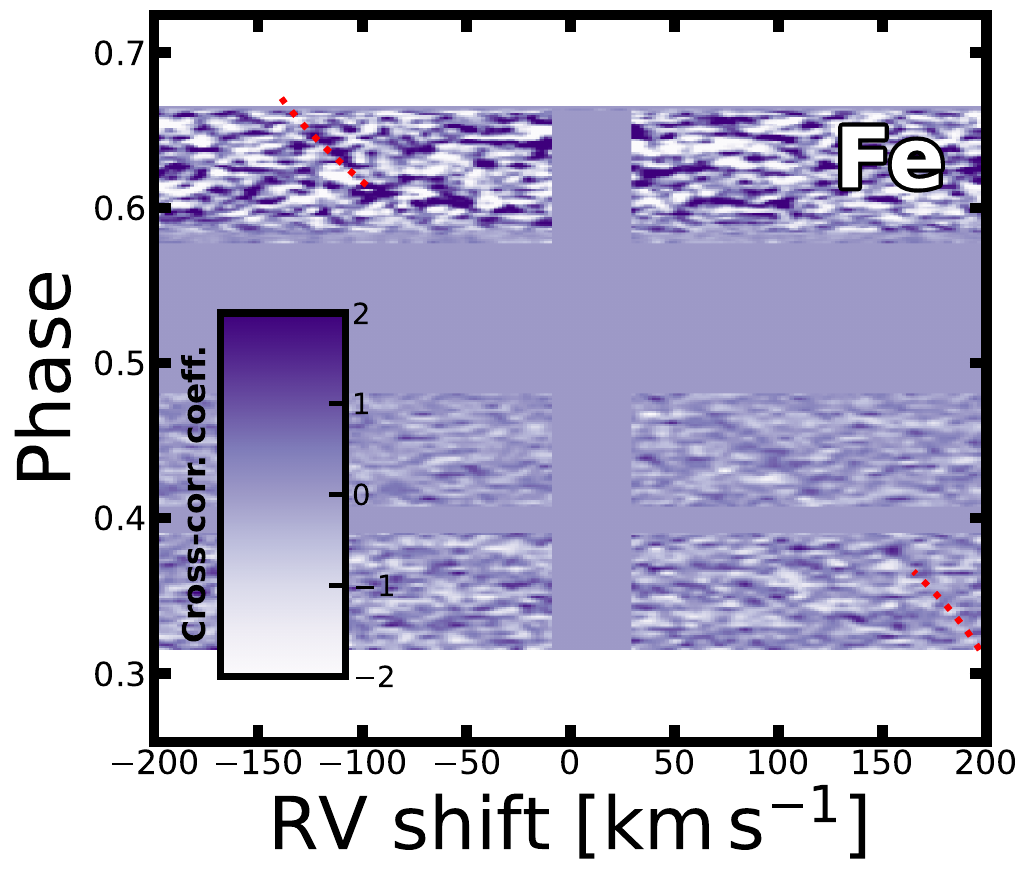}
\hfill
\includegraphics[width=0.47\textwidth]{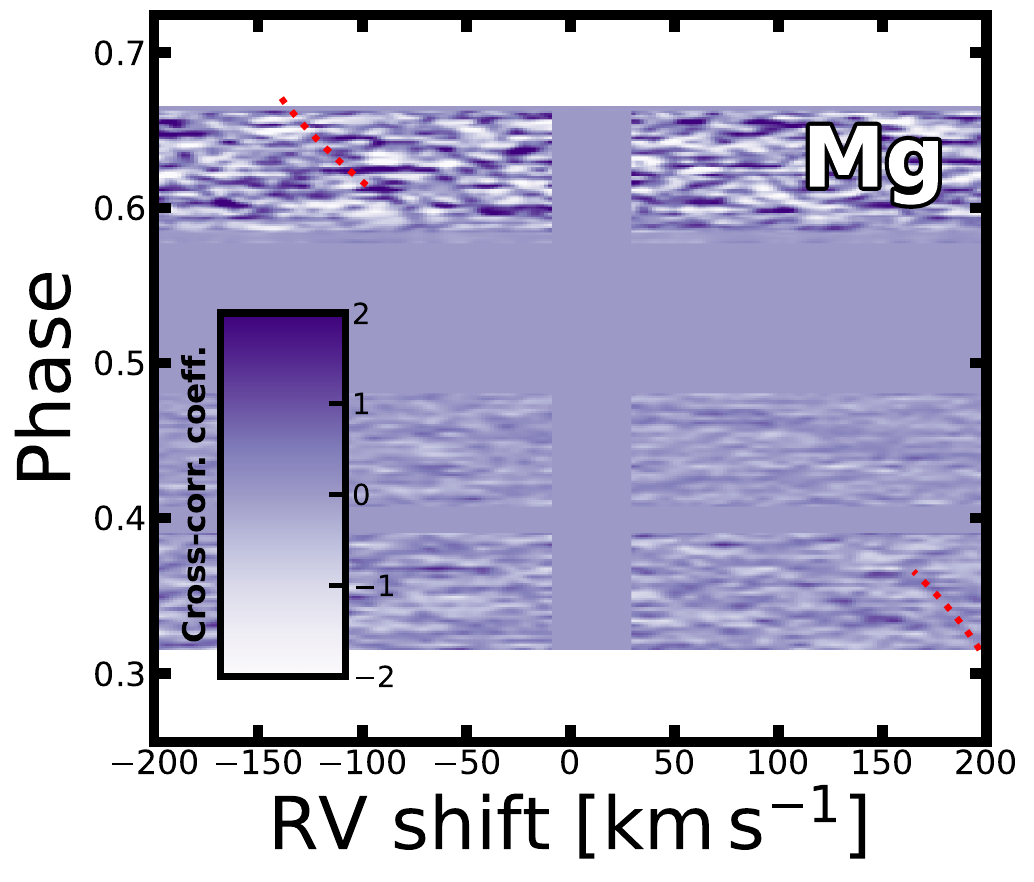}
\vspace{0.cm}
\includegraphics[width=0.47\textwidth]{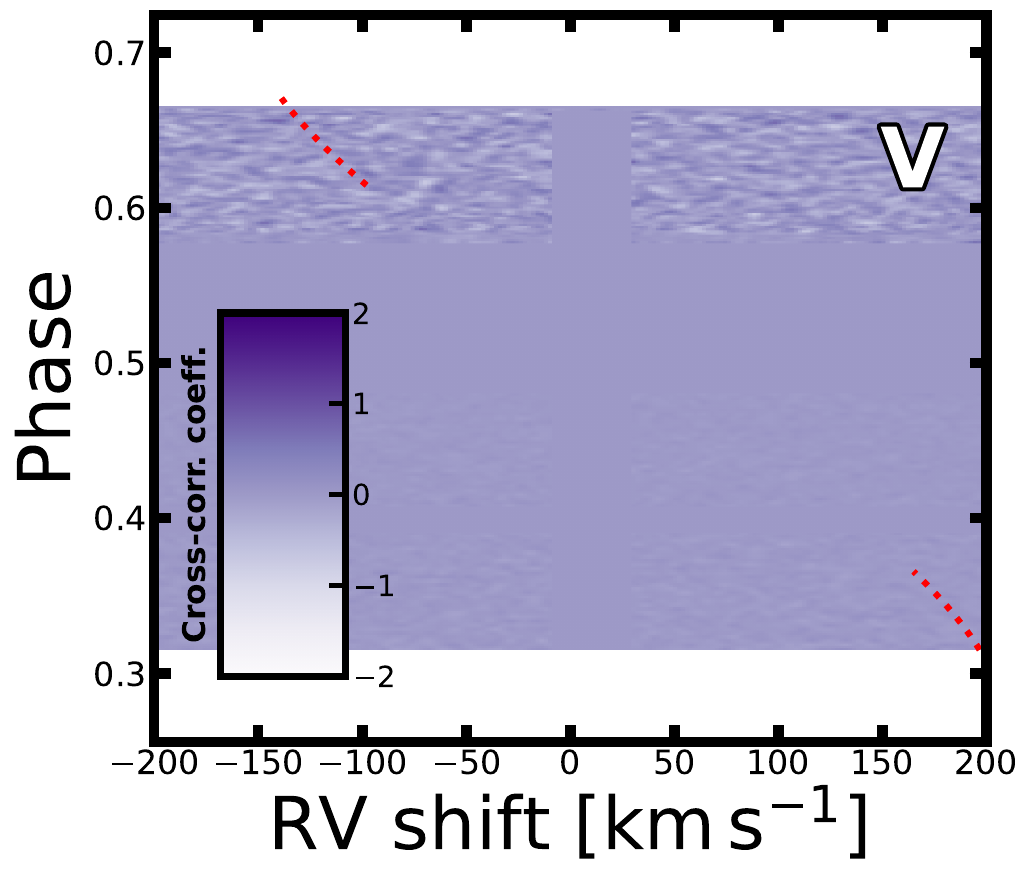}
\hfill
\includegraphics[width=0.47\textwidth]{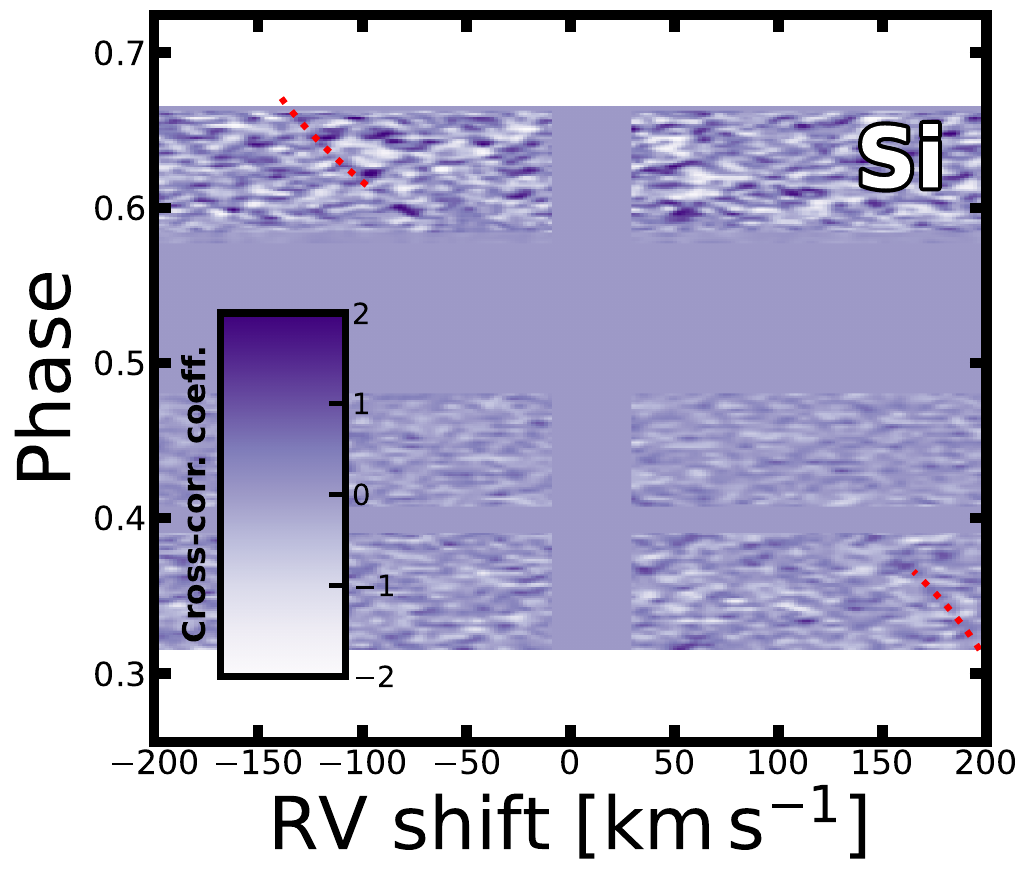}
\vspace{0.cm}
\includegraphics[width=0.47\textwidth]{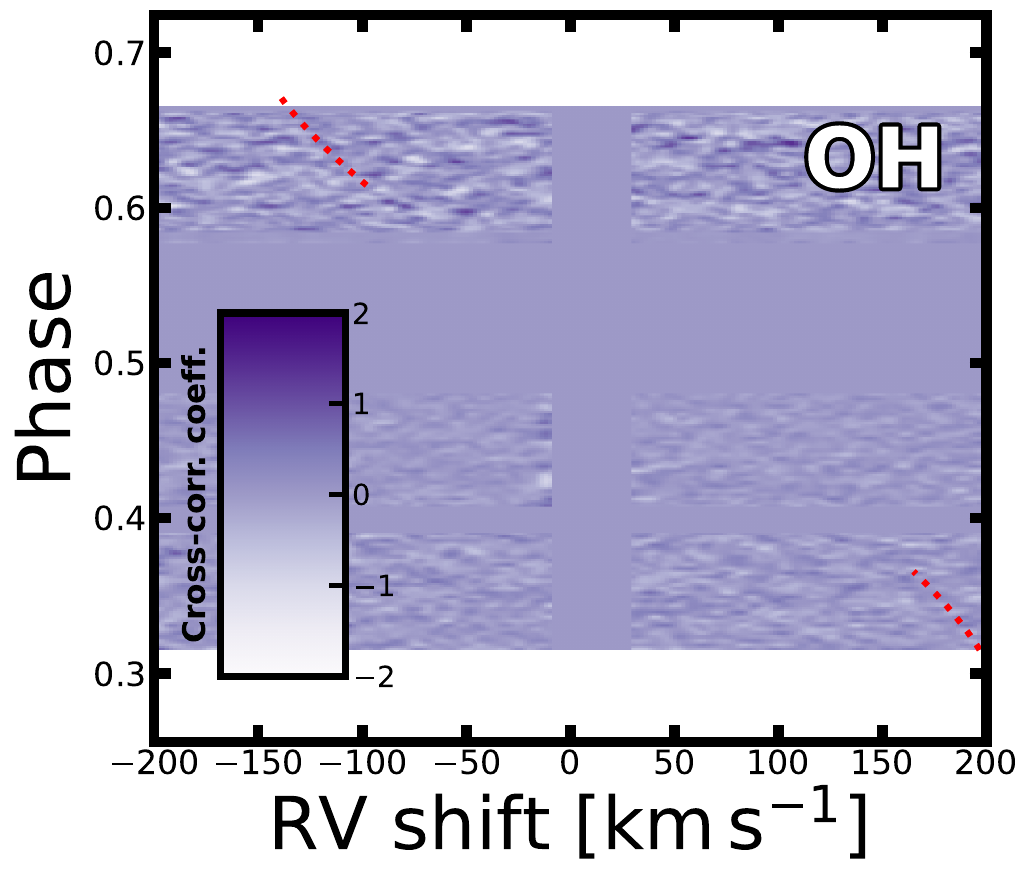}
\caption{Cross-correlation trail maps computed for the ultra-hot Jupiter HAT-P-70b. These maps based on all six NIRPS thermal dayside observations combined together. The red dashed line represents the Keplerian curve of the planet. The colorbar indicates the cross-correlation coefficients. The vertical banded region centred at 10$\pm$20 km~s$^{-1}$ masks residual OH lines from the telluric correction. The horizontal banded region between phases 0.5 and 0.6 approximately corresponds to the orbital segment when the planet is eclipsed by its star. Phases around 0.4 are missing due to technical issues during one of the observations. For details about the observations and the cross-correlation calculation refer to Sect. \ref{sec:Observations} and Sect. \ref{sec:CrossCorrelationAnalysis}.}
\label{Figure:CCFTrailMaps}
\end{figure*}

\clearpage
\section{Supplementary figures: Cross-correlation blob maps}
\begin{figure}[h]
    \centering
    \includegraphics[width=0.48\textwidth]{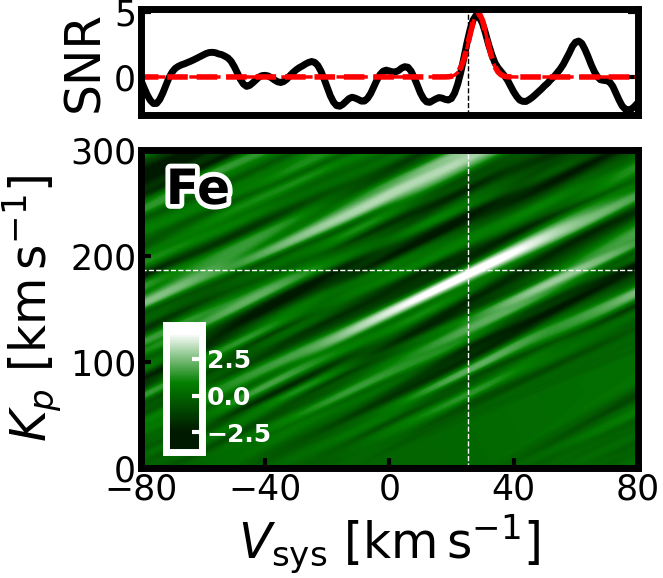}
    \includegraphics[width=0.48\textwidth]{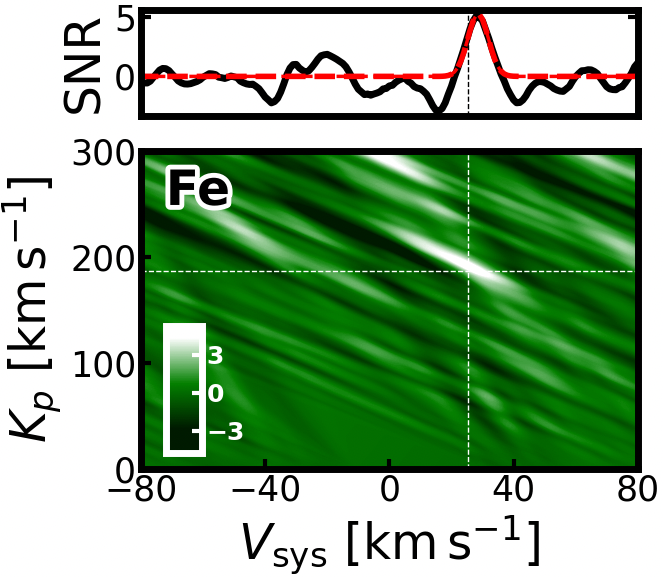}
    \caption{Cross-correlation blob maps for Fe detection in the thermal dayside atmosphere of HAT-P-70b. The purpose is to show the different contribution given by the western dayside (left panel) and the eastern dayside observations (right panel).}
    \label{Figure:Fe_Western_Eastern_Contributions}
\end{figure}
\begin{figure}[h]
    \centering
    \includegraphics[width=0.48\textwidth]{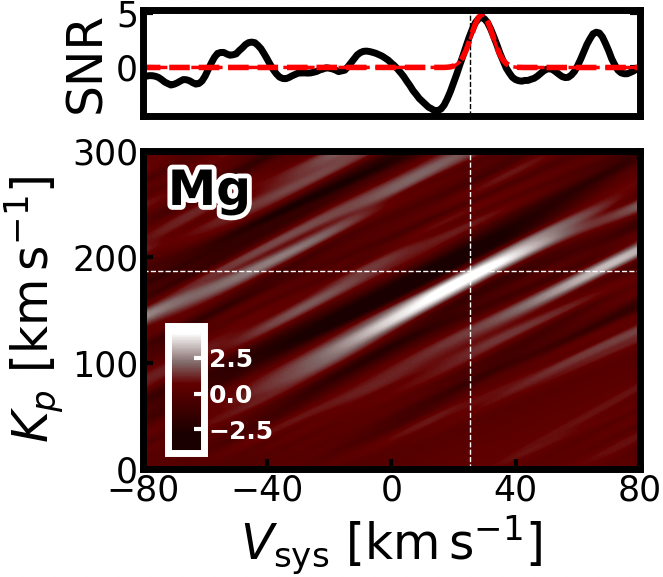}
    \includegraphics[width=0.48\textwidth]{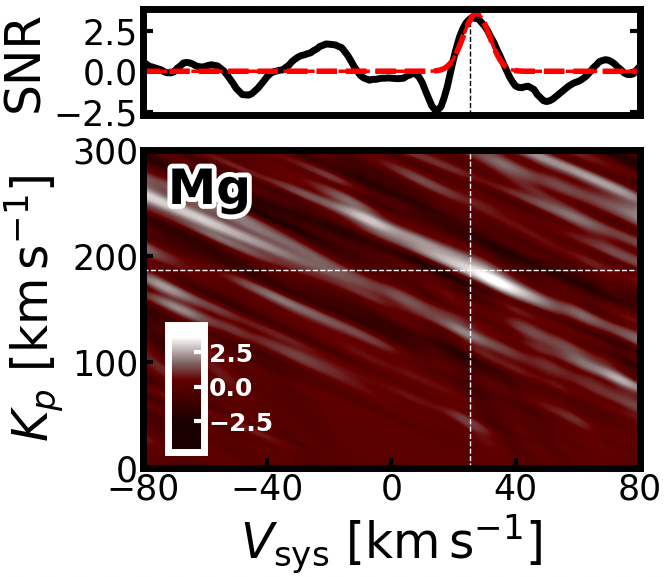}
    \caption{Cross-correlation blob maps for Mg detection in the thermal dayside atmosphere of HAT-P-70b. The purpose is to show the different contribution given by the western dayside (left panel) and the eastern dayside observations (right panel).}
    \label{Figure:Mg_Western_Eastern_Contributions}
\end{figure}
\begin{figure}[h]
    \centering
    \includegraphics[width=0.48\textwidth]{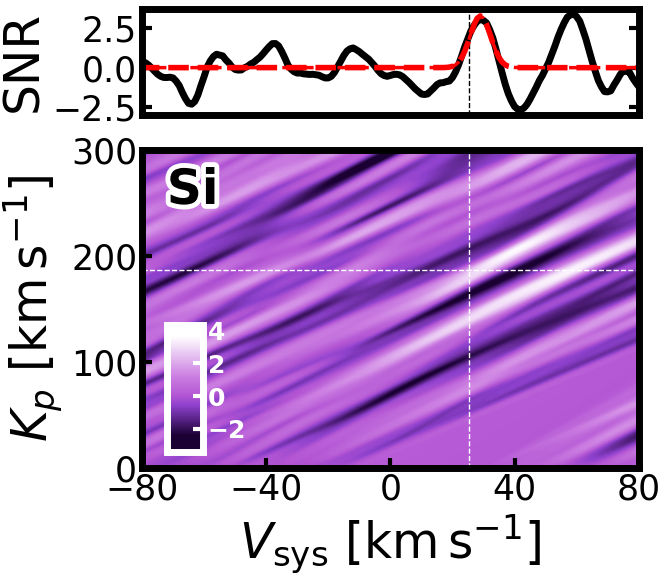}
    \includegraphics[width=0.48\textwidth]{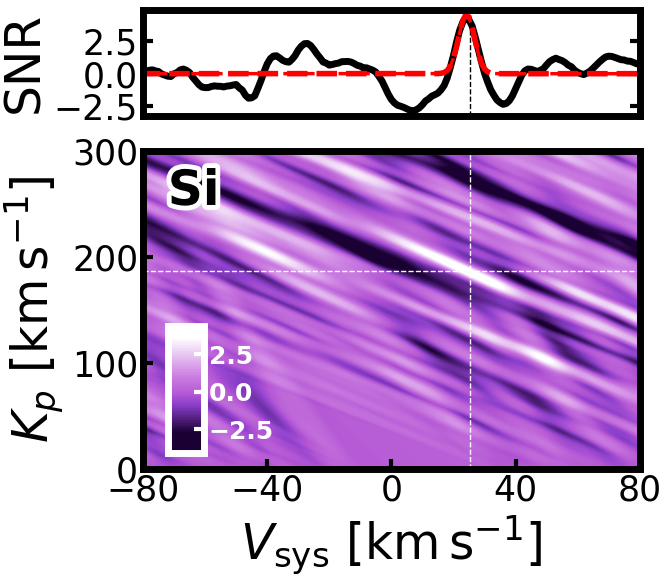}
    \caption{Cross-correlation blob maps for Si detection in the thermal dayside atmosphere of HAT-P-70b. The purpose is to show the different contribution given by the western dayside (left panel) and the eastern dayside observations (right panel).}
    \label{Figure:Si_Western_Eastern_Contributions}
\end{figure}

\clearpage
\section{Supplementary figures: Atmospheric retrieval}
\begin{figure}[h]
    \centering
    \includegraphics[width=0.5\linewidth]{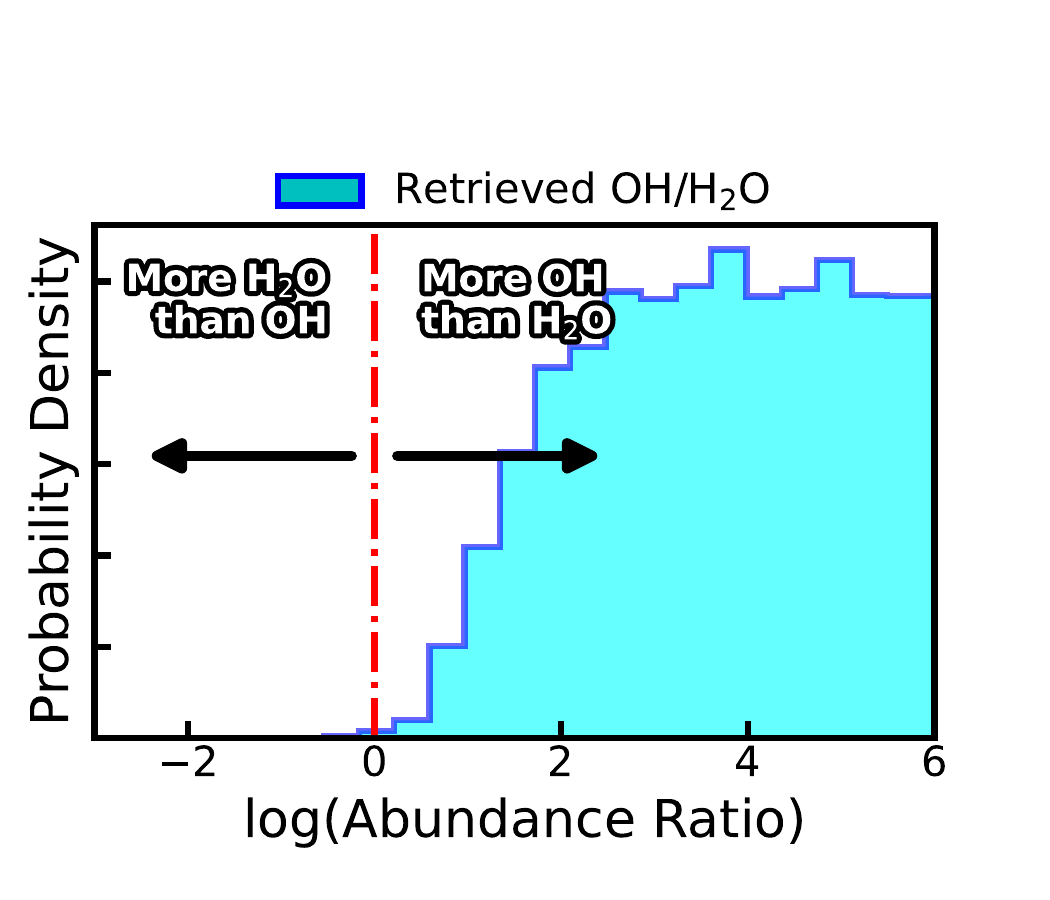}
    \caption{Retrieved $\log$ H$_2$O-to-OH ratio (blue) of the dayside atmosphere of HAT-P-70b from the free retrieval. The 1:1 ratio is indicated with a dotted-dashed red line in order to distinguish the regimes where OH is more abundant than water and viceversa. 
    }
    \label{Figure:OHoverH2O}
\end{figure}
\begin{figure}[h!]
    \centering
    \includegraphics[width=1.
    \linewidth]{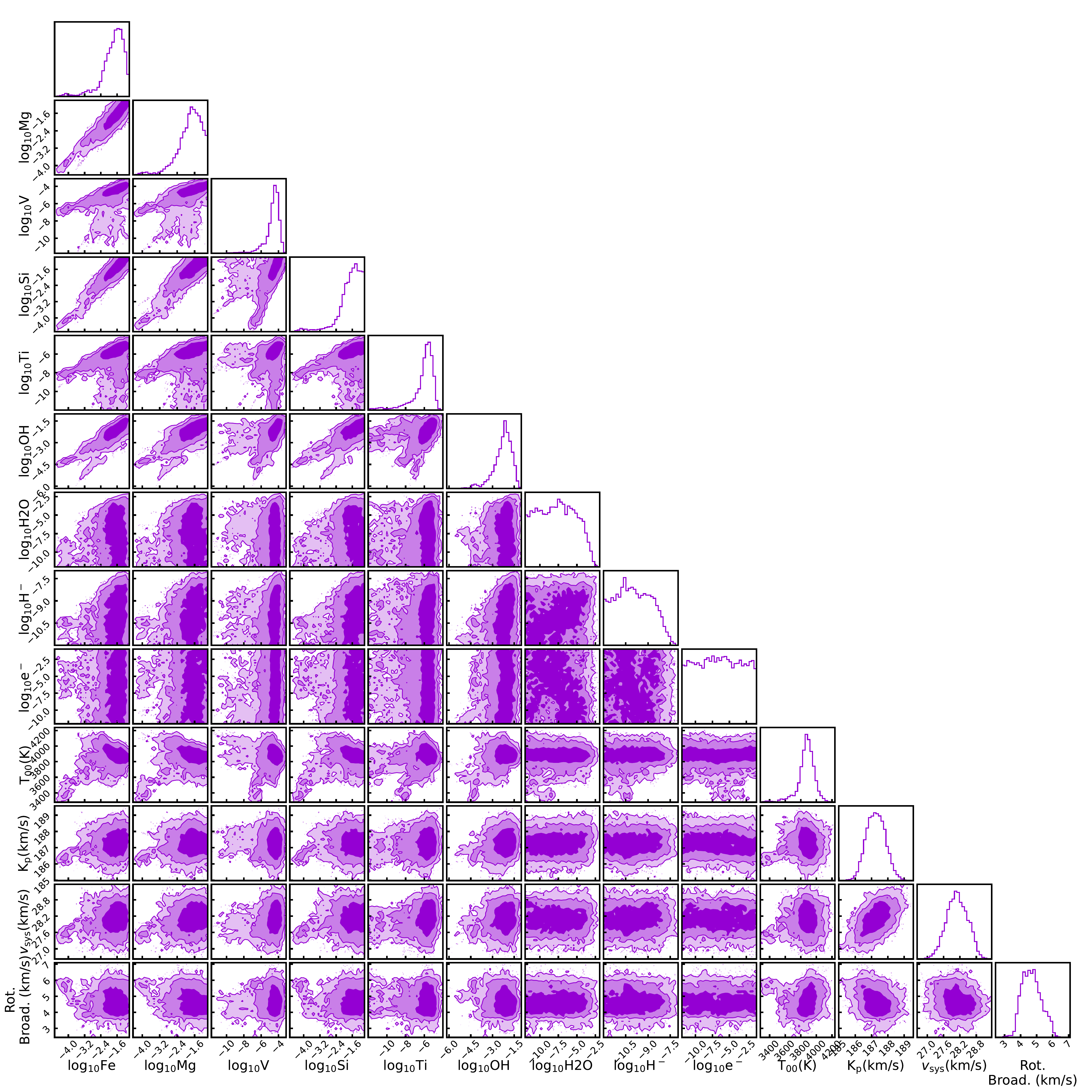}
    \caption{Retrieved constraints on the individual elemental abundance, temperature, and orbital properties of the dayside atmosphere of the ultra-hot Jupiter HAT-P-70b. The corner plot summarise the posterior distribution of the parameters of interest, outcome of the free atmospheric retrieval (as detailed in Sect. \ref{sec:AtmosphericRetrieval}).}
    \label{Fig:CornerPlot}
\end{figure}

\clearpage
\section{Acknowledgments}
\label{sec:Acknowledgments}
\begin{acknowledgements}
The authors thank the anonymous referee for providing insights that improved the quality of this manuscript. This works is based on observations collected at the European Southern Observatory under ESO programmes 112.25P3 and 116.298Z. This project has been carried out within the framework of the National Centre of Competence in Research PlanetS supported by the Swiss National Science Foundation under grant 51NF40\_205606. The authors acknowledge the financial support of the SNSF.
This work has been carried out within the framework of the NCCR PlanetS supported by the Swiss National Science Foundation under grants 51NF40\_182901 and 51NF40\_205606.
DE  acknowledge support from the Swiss National Science Foundation for project 200021\_200726. 
RAl  acknowledges the Swiss National Science Foundation (SNSF) support under the Post-Doc Mobility grant P500PT\_222212 and the support of the Institut Trottier de Recherche sur les Exoplan\`etes (IREx).
RAl, LMa, \'EA, FBa, BB, NJC, RD, CC \& JPW  acknowledge the financial support of the FRQ-NT through the Centre de recherche en astrophysique du Qu\'ebec as well as the support from the Trottier Family Foundation and the Trottier Institute for Research on Exoplanets.
FG  acknowledges support from the Fonds de recherche du Qu\'ebec (FRQ) - Secteur Nature et technologies under file \#350366.
YGCF acknowledges funding from the Swiss National Science Foundation (SNSF) through project P500PN\_217951.
LMa, \'EA, FBa \& RD  acknowledges support from Canada Foundation for Innovation (CFI) program, the Universit\'e de Montr\'eal and Universit\'e Laval, the Canada Economic Development (CED) program and the Ministere of Economy, Innovation and Energy (MEIE).
SCB, ED-M \& NCS  acknowledge the support from FCT - Funda\c{c}\~ao para a Ci\^encia e a Tecnologia through national funds by these grants: UIDB/04434/2020, UIDP/04434/2020.
SCB   acknowledges the support from Funda\c{c}\~ao para a Ci\^encia e Tecnologia (FCT) in the form of a work contract through the Scientific Employment Incentive program with reference 2023.06687.CEECIND and DOI \href{https://doi.org/10.54499/2023.06687.CEECIND/CP2839/CT0002}
{\textcolor{black}{10.54499/2023.06687.CEECIND/CP2839/CT0002.}}
XB, XD \& VY  acknowledge funding from the French ANR under contract number ANR\-24\-CE49\-3397 (ORVET), and the French National Research Agency in the framework of the Investissements d'Avenir program (ANR-15-IDEX-02), through the funding of the ``Origin of Life" project of the Grenoble-Alpes University.
Research activities of the Board of Observational and Instrumental Astronomy at the Federal University of Rio Grande do Norte (NAOS) are supported by continuous grants from the Brazilian funding agency CNPq. This study was financed in part by the Coordena\c{c}\~ao de Aperfei\c{c}oamento de Pessoal de N\'ivel Superior -- Brasil (CAPES) -- Finance Code 001, and by the program CAPES/Print.
BLCM  acknowledge CAPES postdoctoral fellowships and CNPq research fellowships (Grant No. 305804/2022-7).
RC  acknowledges support from the Canada Research Chairs Program and the Natural Sciences and Engineering Research Council of Canada (NSERC).
NBC  acknowledges support from an NSERC Discovery Grant, a Canada Research Chair, and an Arthur B. McDonald Fellowship, and thanks the Trottier Space Institute for its financial support and dynamic intellectual environment.
JRM  acknowledges CNPq research fellowships (Grant No. 308928/2019-9).
ED-M  further acknowledges the support from FCT through Stimulus FCT contract 2021.01294.CEECIND and the support by the Ram\'on y Cajal contract RyC2022-035854-I funded by MICIU/AEI/10.13039/501100011033 and by ESF+.
JIGH, RR, ASM, NN \& AKS  acknowledge financial support from the Spanish Ministry of Science, Innovation and Universities (MICIU) projects PID2020-117493GB-I00 and PID2023-149982NB-I00.
ICL  acknowledges CNPq research fellowships (Grant No. 313103/2022-4).
CMo  acknowledges the funding from the Swiss National Science Foundation under grant 200021\_204847 “PlanetsInTime”.
Co-funded by the European Union (ERC, FIERCE, 101052347). Views and opinions expressed are however those of the author(s) only and do not necessarily reflect those of the European Union or the European Research Council. Neither the European Union nor the granting authority can be held responsible for them.
ASM   acknowledges financial support from the Spanish Ministry of Science and Innovation (MICINN) under the 2024 Ram\'on y Cajal program MICINN RYC2024-050707-I.
ED  acknowledges support from a Banting Postdoctoral Fellowship - NSERC and from the Faculty of Science at the University of Waterloo, and support from the Waterloo Centre for Astrophysics.
KAM  acknowledges support from the Swiss National Science Foundation (SNSF) under the Postdoc Mobility grant P500PT\_230225.
NN  acknowledges financial support by Light Bridges S.L, Las Palmas de Gran Canaria for the Doctoral Thesis ``Habitable Earth-like planets with ESPRESSO and NIRPS", in cooperation with the Instituto de Astrof\'isica de Canarias, and the use of Indefeasible Computer Rights (ICR) being commissioned at the ASTRO POC project in the Island of Tenerife, Canary Islands (Spain). The ICR-ASTRONOMY used for his research was provided by Light Bridges in cooperation with Hewlett Packard Enterprise (HPE).
AKS  acknowledges financial support from La Caixa Foundation (ID 100010434) under the grant LCF/BQ/DI23/11990071.
GAW is supported by a Discovery Grant from the Natural Sciences and Engineering Research Council (NSERC) of Canada.
\end{acknowledgements}

\end{appendix}


\end{document}